\documentclass[12pt,english]{article}
\usepackage[T1]{fontenc}
\usepackage[latin9]{inputenc}
\usepackage{babel}
\usepackage[sorting=none,backend=biber]{biblatex}
\usepackage{geometry}
\usepackage{amsmath}
\usepackage{graphicx}
\usepackage{setspace}
\usepackage{float}

\makeatletter
\@ifundefined{date}{}{\date{}}
\makeatother

\begin{document}
\title{A Study of Dense Suspensions Climbing Against Gravity}
\author{Xingjian Hou and Joseph D. Peterson}
\maketitle
\noindent \begin{center}
\emph{DAMTP, Centre for Mathematical Sciences, University of Cambridge,
CB3 0WA, United Kingdom} 
\par\end{center}
\begin{abstract}
Dense suspensions have previously been shown to produce a range of
anomalous and gravity-defying behaviors when subjected to strong vibrations
in the direction of gravity. These behaviors have previously been
interpreted via analogies to inverted pendulums and ratchets, language
that implies an emergent solid-like structure within the fluid. It
is therefore tempting to link these flow instabilities to shear jamming
(SJ), but this is too restrictive since the instabilities can also be observed
in systems that shear thicken but do not shear jam. As an alternative
perspective, we re-frame earlier ideas about ``racheting" as a ``negative
viscosity'' effect, in which the cycle-averaged motion of a vibrated
fluid is oriented opposite to the direction implied by the cycle-averaged
stresses. Using ideas from the Wyart and Cates modeling framework,
we predict that such a ``negative viscosity" can be achieved in
shear flows driven by oscillating stress with both square and sinuosoidal
wave forms. We extend this same modeling approach to study falling
films in a vibrating gravitational field, where we similarly find
it is possible to attain an overall flow opposite the direction of
gravity. Preliminary experimental findings are also provided in support
of the modeling work. 
\end{abstract}
\global\long\def\ppt{\frac{\partial}{\partial t}}%
\global\long\def\ppy{\frac{\partial}{\partial y}}%
\global\long\def\ub{\boldsymbol{u}}%
\global\long\def\sxy{\sigma_{xy}}%
\global\long\def\sstar{\sigma^{*}}%
\global\long\def\so{\sigma_{0}}%
\global\long\def\dsso{\Delta\sigma/\sigma_{0}}%
\global\long\def\gbar{\bar{\dot{\gamma}}}%

\section{Introduction}

A dense suspension comprises solid particles suspended in fluid, with
particle density high enough that the rheological properties are dominated
by the system's proximity to a ``jamming'' transition.

Suspensions of solid particles in a viscious fluid are found in applications
ranging from construction to comestibles . The ubiquity of so-called
``dense suspensions'' can be partly explained by the difficulties
encountered in transporting dry powders - dense suspensions do not
generate dust, they do not cake, and they are probably less susceptible
to flow-induced segregation effects \cite{ottino2000mixing}. Besides
their relative ease of processing, high density particulate loading
can also improve the viscosity/texture of fluid products like paint
and toothpaste. 

In dense suspensions, there exists a theoretical maximum density at
which particles can be packed together without ``jamming'' the fluid
and obstructing flow. Proximity to jamming can be varied simply by changing particle density,
but under steady flow one must also consider different modes of jamming and changes in
the type of particle-particle contacts present within the system \cite{Clavaud2017Revealing, Poon2018Constraint}.
In particular, high stresses can induce a transition from ``sliding
contacts'' to ``frictional contacts'' \cite{Singh2017Microstructural,Lin2015Hydrodynamic,mari2014shear},
where the latter has fewer degrees of freedom and can therefore produce jammed structures at lower particle densities \cite{Singh2020Shear, wyart2014discontinuous}. The viscosity diverges as constraints are added and degrees of freedom
are removed, in some cases culminating with a "jamming" transition in which the
fluid becomes rigid in response to continued strain in the same direction
 \cite{Cates1998Jamming,Seto2019Shear}. In recent years, there has
been growing interest in tuning/controling the rheological properties
of a dense suspension by applying vibrations that manipulate the relative
proportion of sliding and frictional contacts \cite{ness2018shaken,Lin10774,Gillissen2020Constitutive}.
In these studies, vibrations are always introduced as a shear flow orthogonal to
the principal flow direction.

Vibrations parallel to flow have received comparatively little attention,
except perhaps indirectly in the context of vertically vibrated dense
suspensions (VVDS). Studies of VVDS apply vibrations parallel to gravity,
often yielding dramatic results that defy ordinary expectations. Early
experiments showed that persistent holes can occur at the surface
of VVDS \cite{Merkt2004Persistent}, as though the fluid contained
a ring of elastic material propping the hole open. With increasing
acceleration the holes delocalize and more complex structures emerge
\cite{Merkt2004Persistent}, and subsequent studies have classified
a range of gravity-defying structures including rivers, fingers, and
jumping liquids \cite{vonKann2014Phase}. Shinbrot et al \cite{Shinbrot2015Paradoxical}
have also studied a similar problem, in which dense suspensions showed
apparent ``climbing'' behaviors in the presence of a vertically
vibrating probe.

In all the aforementioned phenomena, VVDS were observed to move up/maintain
shapes against the influence of gravity. As unusual as these observations may seen, motion against gravity has
some precedent in both rigid body mechanics and fluid mechanics, which
we summarize briefly here.

In classical mechanics, there is some precedent for vertical vibrations
favoring motion against gravity: most notably, the inverted vibrating
pendulum \cite{Stephenson1908Stability} and simple racheting mechanisms. Some
 have attempted to extend this analogy for interpreting the behavior
of VVDS \cite{Ramachandran2014Vibro-levitation}, but this explanation may be somewhat
limited as it fails to account for the fundamentally fluid-like character of the
material.

For Newtonian droplets, there is also precedent for vertical vibrations
favoring motion against gravity \cite{BrunetVibration2007}, where
an interplay between inertial and capillary forces allows isolated
droplets to climb up a \emph{partially inclined} surface. However,
this climbing mechanism does not apply to behaviors seen in bulk VVDS,
nor does it permit climbing on surfaces fully inclined to $90^{\circ}$.
We also mention the usual Faraday wave instability, where a quiescent
surface develops a standing wave pattern, but does not exhibit any continuous
climbing behavior \cite{douady1990experimental, Merkt2004Persistent}.

For ordinary shear thickening fluids, Deegan \cite{Deegan2010Stress}
explained climbing behaviors in terms of discontinuous shear thickening
and stress hysteresis in the underlying flow curve. Later simulation
work by Shinbrot et al \cite{Shinbrot2015Paradoxical} proposed that a
 ``racheting mechanism'' might occur for fluids climbing up a vertically
vibrating probe, using similar arguments as Deegan but with a less mathematically
precise framing. However, the simulations by Shinbrot et.
al. only required continuous shear thickening, which suggests a separate climbing
mechanism independent of stress hysteresis. Comparing the ideas from
Deegan et. al. and Shinbrot et. al. we suggest that the source of
vibrations (vibrating platform vs vibrating external probe) may play
a more important role than one might initially expect.

A range of ``climbing'' behaviors can also be found in shear flows
between concentric cylinders, especially where there are Reynolds
stresses (convective acceleration) \cite{gibson1991liquid,Zarraga2001Normal} ,
or elastic stresses (rod climbing) \cite{Bonn2004Rod,hoffman1973determination}
at work. Climbing behaviors such as these are typically explained
in terms of normal-stress differences and/or curved streamlines, and
are generally well understood. The mechanisms for climbing in VVDS
mentioned earlier have so-far excluded any discussion on curved streamlines
and normal stress differences, but this may be an interesting subject
for future inquiry.

For a general explanation of climbing behaviors in VVDS, the stress
hysteresis mechanism proposed by Deegan appears to be most promising,
and in this paper we provide follow-up analysis to (1) explicitly
connect the hysteresis mechanism to changes in inter-particle contacts
and (2) generalize the ``racheting'' idea to a more generic notion
of ``negative apparent viscosity''. These ideas will be supported
by calculations from the Wyart and Cates (WC) modeling framework and
new experimental evidence of sustained ``climbing'' behavior in
VVDS.

The organization of our paper is as follows: Section 2 outlines the
methods of our study, including both the governing equations for the
Wyart-Cates (WC) theory \cite{wyart2014discontinuous} (section \ref{subsec:Theory})
and describes the setup of a proof-of-principle experiment (section
\ref{subsec:Experiments}). Section 3 presents calculations, using
the WC model, for average shear rate under oscillating shear stress
(sections \ref{subsec:Square-wave-stress} and \ref{subsec:Sine-wave-stress})
and average flow rate in a falling film under oscillating gravity
(sections \ref{subsec:square_gravity} and \ref{subsec:sine_gravity}).
We also provide a limited discussion on the effect of finite inertia
(section \ref{subsec:intertia}). Section 4 describes the results
of our experimental efforts and provides an intrepretation of those
results following the modeling work of section
\ref{sec:Modeling-Results}. Section 5 summarizes the results of our
study and suggests directions for future research.

\section{\label{sec:Methods}Methods}

In this paper, we aim to demonstrate how the inducement of frictional
contact in both discontinuous shear thickening (DST) and shear jamming (SJ) suspensions can help
explain observations of gravity-defying flow in vertically-vibrated
dense suspensions (VVDS). To achieve this, we require a modeling tool
that (1) relates rheology to fluid microstructure, (2) is large in
scale compared to particle size (i.e. continuum models), (3) is suitable
for reversing flows, (4) is suitable for spatially-resolved flows,
and (5) has minimal complexity. It is our view that no exising modelling
approach will satisfy all of these demands, and so we defer to WC
as a minimal model for preliminary study with acknowledged weaknesses
in criterion (3) and (4). These weaknesses are discussed in more detail
at the end of section \ref{subsec:Theory}.

In our experimental efforts, we will aim to conduct proof-of-principle
tests for the basic model predictions, and also provide some new experimental
results that more unabiguously isolate the ``climbing'' behavior
of VVDS. Future experimental efforts in this vein will require more
precise equipment for a systematic study.

The subsections that follow provide a more complete description of
the modeling tools and experimental setup that support the results
in sections \ref{sec:Modeling-Results} and \ref{sec:Experimental-Results},
respectively.

\subsection{\label{subsec:Theory}Theory}

To qualitatively capture continuous/discontinuous shear thickening
phenomena in dense suspensions, we will leverage modeling tools previously
introduced by Wyart and Cates (WC) \cite{wyart2014discontinuous}.
To adapt WC to time-dependent flows, we simply exchange the steady
shear rate for our time-dependent shear rate.

The physical picture underlying the WC model is summarized as follows:
at low stresses, particles are always separated by a thin lubrication
layer that allows for sliding contacts. Above a critical stress, however,
lubrication films begin to collapse and particles experience a frictional
contact that permits rolling but restricts sliding. With increasing
stress, more contacts become frictional and the contact network becomes
more constrained in its available movements, leading to shear thickening
at the bulk scale. This physical picture is captured mathematically
by the WC model equations below:

\begin{equation}
\begin{aligned}\sigma & =\eta\dot{\gamma},\eta=\frac{\eta_{0}}{(\phi-\phi_{J})^{2}},\\
\phi_{J} & =f\phi_{R}+(1-f)\phi_{S},\\
f & =\textup{exp}(-\frac{\sigma^{*}}{\Pi})
\end{aligned}
\label{eq:WC}
\end{equation}

where $\dot{\gamma}$ is the shear rate of the suspension, $\sigma$
is the shear stress on the suspension, $\Pi=\textup{tr}(\boldsymbol{\Sigma}/3)$
is the particle pressure with $\boldsymbol{\Sigma}$ being the particle
stress tensor (note that $\sigma$ denotes the $xy$ component of
$\boldsymbol{\Sigma}$ in our work). For steady simple shear flow, the particle
pressure $\Pi=\text{tr(\ensuremath{\boldsymbol{\Sigma}/3)}}$should
be positive and proportional to the shear stress $\Pi\propto|\sigma|$,
and for our purposes the constant of proportionality can be absorbed
into the choice of $\sigma^{*}$, which is the critical stress needed
to drive particles into contact. For a given fraction of frictional
contacts, $f$, Eq. \ref{eq:WC} prescribes a jamming fraction $\phi_{J}$
that interpolates linearly between the jamming fraction for purely
sliding constraints, $\phi_{S}$, and purely rolling constraints,
$\phi_{R}$. Note that $f$ is a function of the scaled particle pressure
$|\sigma|/\sigma^{*},$ with frictional contacts being favored for
large particle pressures. For the purpose of this paper, we will assume
$\phi_{R}=0.57,\phi_{S}=0.64$. The viscosity of the suspension $\eta$
is related to the actual volume fraction of the suspension $\phi$.
The pre-factor in the Krieger--Dougherty equation \cite{Krieger1959}
is denoted by $\eta_{0}$ and is proportional to the viscosity of
the suspending fluid.

\begin{figure}[H]
\begin{centering}
\includegraphics[width=0.65\columnwidth]{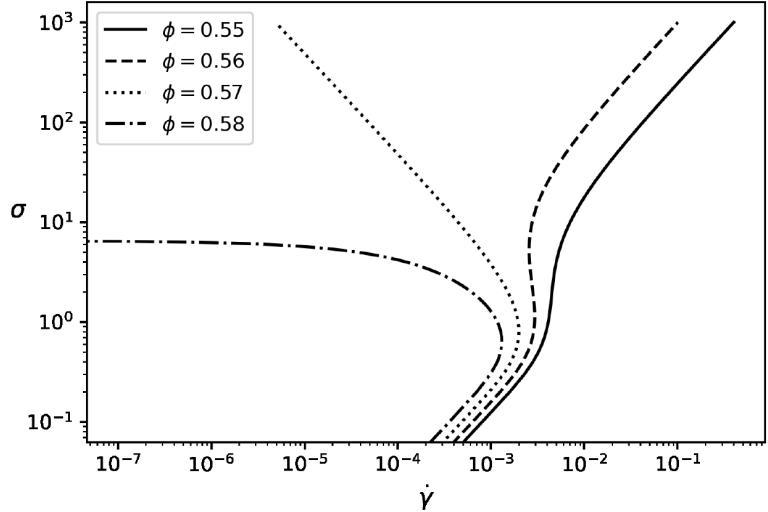}
\par\end{centering}
\caption{\label{fig:WC_steady} Predictions from the Wyart and Cates (WC) model
(Eq. \ref{eq:WC}) for the steady shear stress $\sigma$ as a function
of shear rate $\dot{\gamma}$ for particle volume fractions $\phi=0.55,0.56,0.57,0.58$,
assuming WC parameters $\phi_{R}=0.57,\phi_{S}=0.64,\sigma^{*}=1,\eta_{0}=1$.}
\end{figure}

The principal success of the WC model is its ability to qualitatively
capture a progression from continuous shear thickening (CST) to discontinuous
shear thickening (DST) and shear jamming (SJ) with increasing particle
volume fraction. In Fig. \ref{fig:WC_steady}, we plot example ``flow
curves'' for the shear stress $\sigma$ as a function of the shear
rate $\dot{\gamma}$ over a range of volume fractions $\phi$ covering
this transition. Given the assumed values $\phi_{R}=0.57,\phi_{S}=0.64$,
the WC model predicts that hysteresis and DST occur in the range $0.555<\phi<0.64$.
For $0.57\leq\phi<0.64$, it is possible for SJ to occur, i.e. obtaining
$\dot{\gamma}=0$, once $\sigma$ exceeds a critical value. For $\phi<0.555$,
only CST is observed.

Where the shear stress is known, Eq. \ref{eq:WC}, can be used to
determine the corresponding shear rate:

\begin{equation}
\dot{\gamma}=\begin{cases}
\frac{\sigma}{\eta_{0}}\left[(\phi_{R}-\phi_{S})\textup{exp}(-\frac{\sigma^{*}}{|\sigma|})+(\phi_{S}-\text{\ensuremath{\phi)}}\right]^{2}, & \phi<\phi_{J}\\
0, & \phi>\phi_{J}
\end{cases}\label{eq:gdot_WC}
\end{equation}

Eq. \ref{eq:gdot_WC} can also be generalized for a known time-dependent
shear stress, $\sigma(t)=\sigma_{0}+\Delta\sigma(t)$. For a forcing
that's periodic on time $T,$we can define an average shear rate $\bar{\dot{\gamma}}$
by:

\begin{equation}
\bar{\dot{\gamma}}=\frac{1}{T}\int_{0}^{T}dt\dot{\gamma}(t)\label{eq:averaging shear rates}
\end{equation}

Likewise, we can define a cycle-averaged viscosity $\bar{\eta}=\sigma_{0}/\bar{\dot{\gamma}}$,
assuming $\Delta\sigma(t)$ has zero mean. For Newtonian fluids -
and indeed for most non-Newtonian fluids, the sign of $\bar{\eta}$
is always positive. However, our calculations in sections \ref{subsec:Square-wave-stress}
and \ref{subsec:Sine-wave-stress} suggest that this may not always
be the case for dense suspensions.

The coupling between fluid motion\footnote{For convenience, equation \ref{eq:NS} is written in the frame of
the vibrating support. In the lab frame the suspension is under constant
gravity $\boldsymbol{g}_{0}$ and the support vibrates parallel to
gravity with acceleration $-\Delta\boldsymbol{g}(t)$ of period $T$,
so the equivalent gravity is $\boldsymbol{g}(t)=\boldsymbol{g}_{0}+\Delta\boldsymbol{g}(t)$
in the vibrating frame.} and fluid stresses in a fluid film is determined by via the Navier-Stokes
equation, conserving momentum:

\begin{equation}
\rho(\frac{\partial\ub}{\partial t}+\ub\cdot\nabla\ub)=-\nabla p+\nabla\cdot\boldsymbol{\sigma}+\rho\boldsymbol{g}\label{eq:NS}
\end{equation}

We also assume an incompressible flow:

\begin{equation}
\nabla\cdot\ub=0
\end{equation}

For uni-direction flow, $\ub=u_{x}\boldsymbol{e}_{x}$, and Eq. (3)
becomes:

\begin{equation}
\rho\frac{\partial u}{\partial t}=\frac{\partial\sigma}{\partial x}+\rho g_{x}\label{eq:1D_NS}
\end{equation}

Eq. \ref{eq:1D_NS} will be applied to model oscillatory shear flows,
$g_{x}=0$, and falling films under oscillating gravity , $g_{x}=g_{0}+\Delta g(t)$,
where $g_{0}$ is normal gravity and $\Delta g(t)$ is a zero-mean
oscillation.

For the remainder of this report, we will conduct our analysis in
terms of non-dimensionalized equations. A set of characteristic scales
is defined to non-dimensionalize Eq. \ref{eq:1D_NS}. The time scale
is $\tau_{C}=T$, the period of oscillation. The length scale is $y_{c}=H$,
the thickness of the fluid film. We assign a stress scale $\sigma_{C}$
on the basis of the average forcing: for flows driven by an oscillating
shear stress $\sigma(t)$ of period $T$, we choose $\sigma_{C}=\sigma_{0}$,
where $\sigma_{0}=\frac{1}{T}\int_{0}^{T}dt\sigma(t)$ is the average
shear stress. For falling film flow driven by oscllating gravity $g_{x}(t)$,
we choose $\sigma_{C}=\sigma_{0}$ where $\sigma_{0}=\rho g_{0}H$
gives the wall shear stress for a Newtonian fluid under fixed gravity
$g_{0}=\frac{1}{T}\int_{0}^{T}dtg_{x}(t)$. Given these characteristic
stresses, we use the zero shear viscosity of the fluid $\eta_{\dot{\gamma}=0}$
to estimate a characteristic velocity, $u_{C}=H\sigma_{C}/\eta_{\dot{\gamma}=0}$.
Rescaling all variables in this way, we obtain:

\begin{equation}
\text{Re}\ppt u=-\frac{g_{x}}{g_{0}}+\frac{\partial\sigma}{\partial x}\label{eq:NS_1D_ndim}
\end{equation}

where:

\begin{equation}
\text{Re}=(\frac{H^{2}\rho}{\eta_{\dot{\gamma}=0}T})\label{eq:Re expression}
\end{equation}

The Reynolds number is a dimensionless variable comparing the typical
scale of inertial forces to viscous forces. For the purpose of study,
we assume Re $\ll1$, as we are interested in high viscosity fluids,
$\eta_{\dot{\gamma}=0}\sim1$Pa$\cdot$s, small lengthscales, $H\sim1$mm,
and modest cycle times, $T\sim0.1$s. The system is also influenced
by (1) the typical stresses in the fluid relative to the stress required
for frictional contact $\sigma_{0}/\sigma^{*}$, where $\sigma_{0}=\sigma_{C}$
for a fluid film; (2) the relative amplitude of the oscillations $\Delta\sigma/\sigma_{0},\Delta g/g_{0}$,
where $\Delta\sigma,\Delta g$ is the amplitude of the oscillations
$\Delta\sigma(t)$ and $\Delta g(t)$.

In the introduction to this section, we noted that there are at least
two major limiations to this modeling approach. To close this section,
we expand upon those limitations in more detail.

First, WC is strictly suited to steady shear flows and cannot account
for time-dependent flows , reversing flows , and it lacks a tensorial
structure for general 3D flows \cite{Cates2021Constitutive, Gillissen2020Constitutive, wyart2014discontinuous}.
Unfortunately, existing models that overcome these limitations have
considerably higher complexity \cite{Gillissen2020Constitutive} and
are still limited in many respects (see next paragraph). In any case,
we feel that a minimal model like WC is appropriate for a preliminary
study, provided one keeps the model's limitations in mind through
the final analysis.

Second, WC is not suited for spatially resolved flows in which one
can encounter particle migration \cite{chacko2018dynamic, boyer2011dense}
and non-local microstructure relaxation processes \cite{henann2014continuum},
neither of which are included in our present approach. These physics
may be of particular relevance where there is an unstable section
of a dense suspension's flow curve \cite{olmsted2008perspectives, saint2018uncovering, rathee2017localized, darbois2020surface},
and this is the portion of the flow curve most relevant to our propsed
mechanism for a ``negative viscosity'' effect. However, to our knowledge there is
not yet an established modeling framework that would address all of these
concerns, and so we once again defer to the simpler WC model.

With respect to both of the limitations above, we are effectively
assuming that the WC model captures essential features of flows with
DST even where it fails to capture the details of how those flows
develop spatially and temporally. This is a severe approximation,
but it is also a reasonable concession given the limitations in the
current space of available modeling tools.

\subsection{\label{subsec:Experiments}Experiments}

\begin{figure}[H]
\begin{centering}
\includegraphics[width=0.6\textwidth]{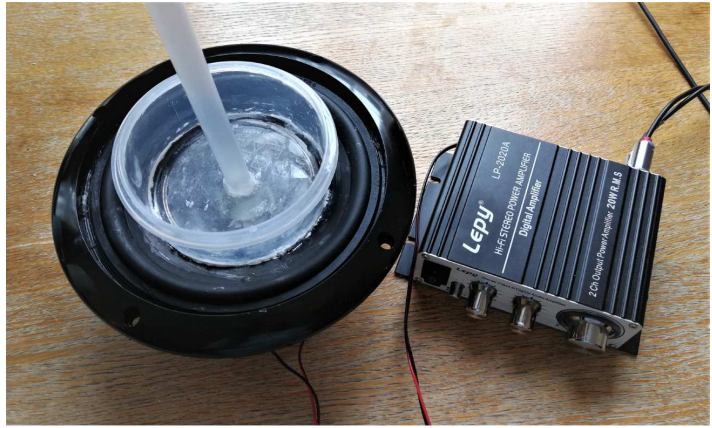}
\par\end{centering}
\caption{\label{fig:Experimental_Setup}Experimental setup: The loudspeaker
and the digital amplifier connected to each other. A plastic container
was adhered to the loudspeaker to contain the fluid. Experiments were
conducted with the plastic container alone or with additional structures
attached, including a plastic straw.}
\end{figure}

In the experiment, an oscillating signal is generated via an online
tone generator \cite{online} connected to a digital amplifier (Lepy
Lp 2020A). A loudspeaker (Pyle PLMR61B, 120W) connects to the amplifier,
providing oscillation. To contain the fluid, we adhered a plastic
container (dimensions 10cm diameter, 4cm height) to the surface of
the loudspeaker using hot-melt adhesive. Experiments were conducted
with the plastic container alone or with additional structures attached,
including (as seen in Fig. \ref{fig:Experimental_Setup}) a plastic
straw (1.2 cm diameter, 20 cm height).

In the experiment, the frequencies of oscillations range between $1\sim25Hz$.
At 10Hz, where many of our experiments were conducted, the maximum peak-to-peak
displacement was observed to be approximately 1cm, corresponding to
a peak acceleration of $\Delta g\approx20m/s^{2}$.

\section{\label{sec:Modeling-Results}Modeling Results}

In this section we will employ Eqs. \ref{eq:WC} and \ref{eq:NS_1D_ndim}
(with Re $=0$) to model the effect of vibrations/oscillations parallel
to the direction of flow, especially as it pertains to a cycle-averaged
shear rate $\gbar$ (oscillating shear flows, sections \ref{subsec:Square-wave-stress}
and \ref{subsec:Sine-wave-stress}) or a cycle-averaged flow rate
(oscillating falling films, sections \ref{subsec:square_gravity}
and \ref{subsec:sine_gravity}). Finally, in section \ref{subsec:intertia},
the effect of finite inertia on the motion of fluid film with sinusoidal
forcing is discussed and it is found that a small amount of inertia
gives a first correction term that averages to zero.

\begin{figure}[H]
(a)%
\begin{minipage}[t]{0.49\columnwidth}%
\includegraphics[scale=0.47]{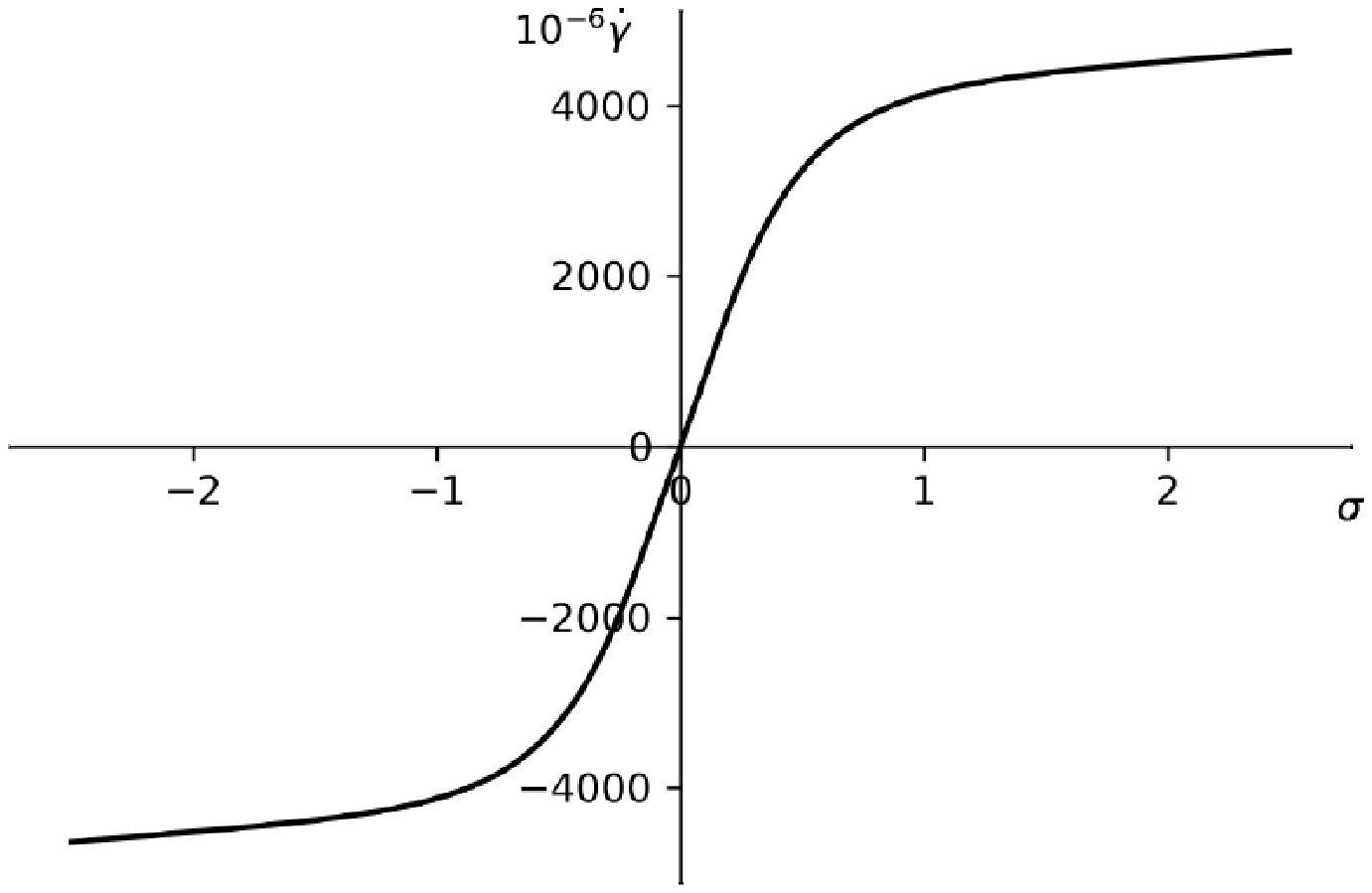}%
\end{minipage}\hfill{}(b)%
\begin{minipage}[t]{0.49\columnwidth}%
\includegraphics[scale=0.47]{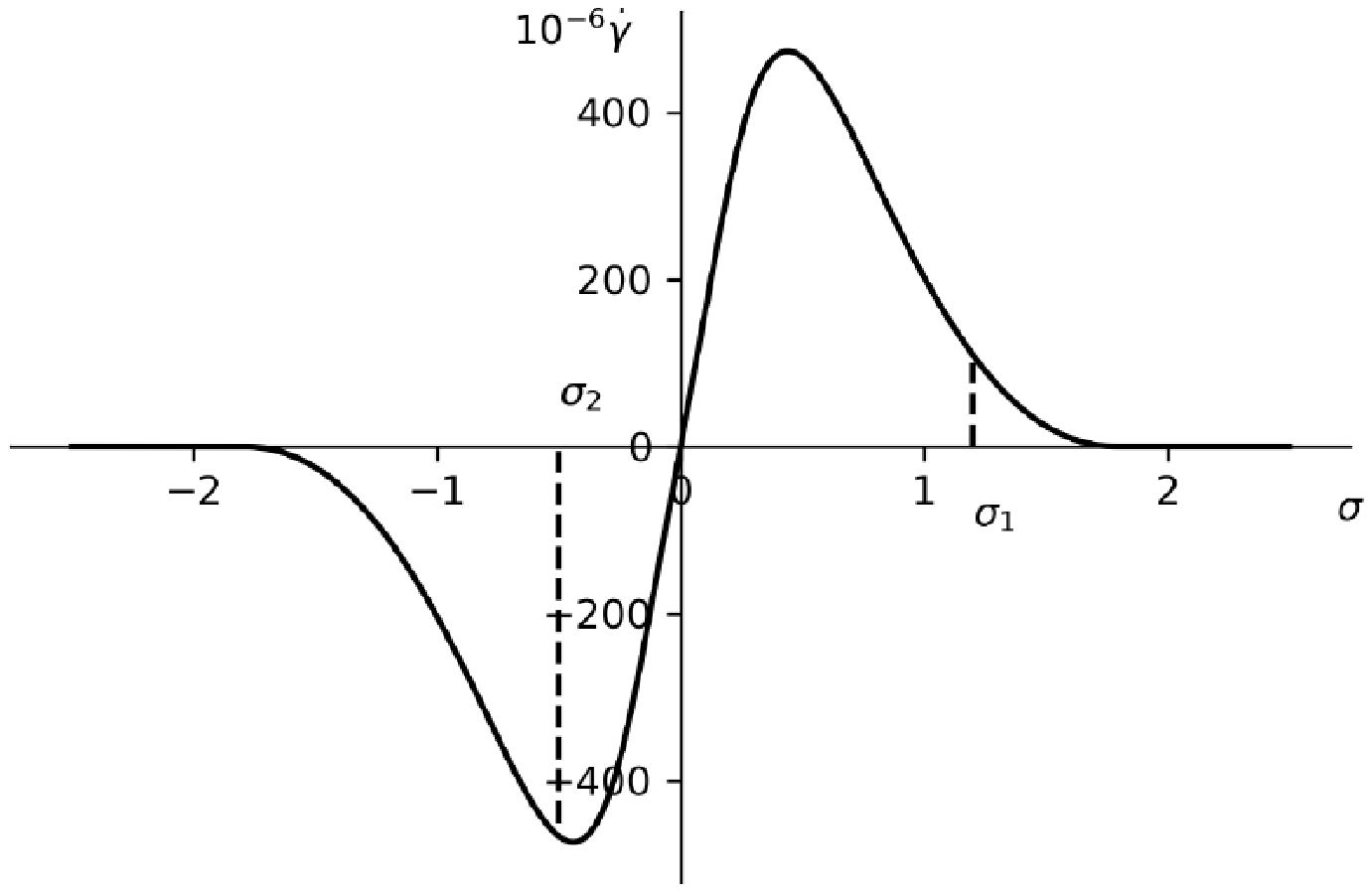}%
\end{minipage}\caption{\label{fig:mechanism}
(a) Shear rate $\dot{gamma}$ - stress $\sigma$
relationship for suspension with $\phi=0.55$. It is impossible to
obtain $\sigma_{1},\sigma_{2}$ such that $\sigma_{1}+\sigma_{2}>0,\dot{\gamma_{1}}+\dot{\gamma_{2}}<0$.
(b) Shear rate $\dot{\gamma}$ - stress $\sigma$ relationship for
suspension with $\phi=0.6$. It is possible to obtain $\sigma_{1},\sigma_{2}$
such that $\sigma_{1}+\sigma_{2}>0,\dot{\gamma_{1}}+\dot{\gamma_{2}}<0$.}
\end{figure}

Before proceeding with detailed calculations, we first provide a general
discussion on the underlying mechanism by which one might obtain an
apparent negative viscosity from a dense suspension. Fig. \ref{fig:mechanism}(a)
gives the flow curve for a sample CST fluid, $\phi=0.55<0.555$. Here,
shear rates always increase with increasing shear stress, and it is
not possible to choose a set of points along the flow curve for which
the average shear rate and the average shear stress do not have the
same sign. However, for a DST fluid like that shown in Fig. \ref{fig:mechanism}(b),
there is a portion of the flow curve where the shear stress is decreasing
with increasing shear rate. Neglecting, for the moment, concerns over
the stability of these portions of the flow curve, it is now possible
to choose a pair of points (e.g. corresponding to the start of the
high viscosity branch at positive shear rates and the end of the low
viscosity branch at negative shear rates) such that the average shear
rate $\bar{\dot{\gamma}}=(\dot{\gamma_{1}}+\dot{\gamma_{2}})/2$ corresponds
to motion opposite direction to average shear stress $\sigma_{0}=(\sigma_{1}+\sigma_{2})/2$,
thus giving the appearance of a ``negative viscosity''.

This proposed mechanism makes no special distinction between SJ and
DST materials - the ``negative viscosity'' effect is made possible
through the existence of a turning point in the underlying flow curve.
How and where a ``negative viscosity'' effect appears may vary depending
on the details of the underlying flow curve and the time-dependent
stress protocol, as we will see in the sections that follow.

Here, we again remind the reader that our modeling approach carries
two limitations with respect to the above proposed mechanism as discussed
in section \ref{subsec:Theory}: namely, the WC model is not really
equipped to model unsteady or spatially resolved flows, as is necessary
for any proper study of the unstable brach of the flow curve \cite{olmsted2008perspectives,chacko2018dynamic, rathee2017localized, saint2018uncovering, darbois2020surface}.
In light of these limitations, we have argued that the use of WC is
justified if one expects that the basic hysteretic structure of the
WC flow curve is more or less preserved even where a uniform flow
is not attainable.

\begin{figure}[H]
\begin{centering}
\includegraphics[scale=0.47]{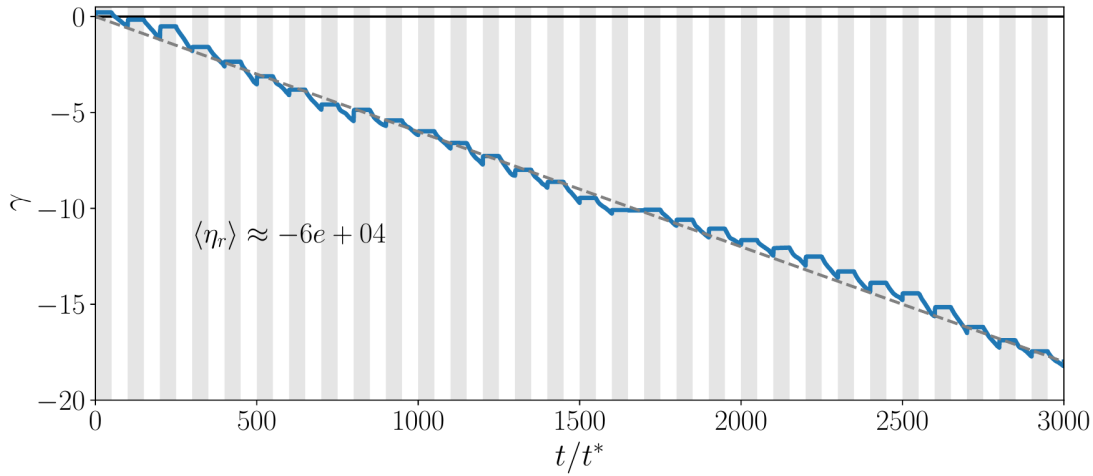}
\par\end{centering}
\caption{\label{fig:particle_sim} Accrued strain as a function of time for
particle-based calculations of the 2D critical load model. The calculations
considered a bi-disperse collection of 100 discs (66 of size 1, 34
of size 1.4) with a friction coefficient of 2, covering an area fraction
of 0.78. The forward and reverse shear are produced via a square-wave
shear stress protocol, alternating 50 units of time at +20 critical
stress units followed by 50 units of time at -1/20 critical stress
units. The negative sloping trend line indicates that strain is occruing
opposite to the average stress, and the slope of the curve provides
an estimate of the apparent viscosity.}
\end{figure}

To partially assuage concerns over the simplicity of our modeling
approach, we assure the reader that 2D particle-based simulations
using the critical load model \cite{mari2014shear} affirm the premise
of our mechanism for a ``negative viscosity'' effect. The results
shown in Figure \ref{fig:particle_sim} were generously contributed by
by Roman Mari. These proof-of-principle calculations consider a bi-disperse
collection of 100 discs (66 of size 1, 34 of size 1.4) with a friction
coefficient of 2, covering an area fraction of 0.78. The forward and
reverse shear are produced via a square-wave shear stress protocol,
alternating 50 units of time at +20 critical stress units followed
by 50 units of time at -1/20 critical stress units. Under large positive
stress, the system jams and ceases to flow such that the small negative
stress dominates the cycle-averaged shear rate, as evidenced by the
negatively-sloping trend line in accrued strain over time.

Thus, while we acknowledge that our choice of constitutive model is
limiting, we are confident that the underlying premise of our proposed
mechanism is basically sound. Follow-up studies seeking a quantitative
confrontation to experimental results or particle-based simulations
will benefit from more detailed constitutive modeling tools, but for
the time being it is our view that the WC model provides the best
balance of microscopic insight and computational simplicity.

As a closing note to this introductory section, our use of the term
``negative viscosity'' should be understood as an interpretative
framework rather than an actual material property. A simple way to
demonstrate the limitations of this interpretive framework is to consider
the possibility of a ``zero viscosity'' effect, where the time-averaged
stress is zero but the time-averaged shear rate is non-zero. Our study
only looks at time-symmetric oscillations, where a zero-average stress
is only possible for a zero-average shear rate, but for time-asymmetric
oscillations a ``zero viscosity'' may indeed be possible. However,
in such cases the directionality of the flow is predetermined by the
asymmetry of the forcing (and unchanged by small perturbations to
the stress) so it is not clear that a ``zero viscosity'' effect
would exhibit any non-trivial behaviors beyond those already discussed
for a ``negative viscosity'' effect.

\subsection{\label{subsec:Square-wave-stress}Square-wave Oscillations in Shear
Stress}

In this section, we will consider flow driven by a defined time-dependent
shear stress following a square-wave variation in time:

\begin{equation}
\sigma(t)=\begin{cases}
\sigma_{0}+\Delta\sigma & \text{if }t\text{ (mod }1)<1/2\\
\sigma_{0}-\Delta\sigma & \text{if }t\text{ (mod }1)\geq1/2
\end{cases}\label{eq:square_stress}
\end{equation}

\begin{figure}[H]
\begin{centering}
(a)\includegraphics[width=0.47\columnwidth]{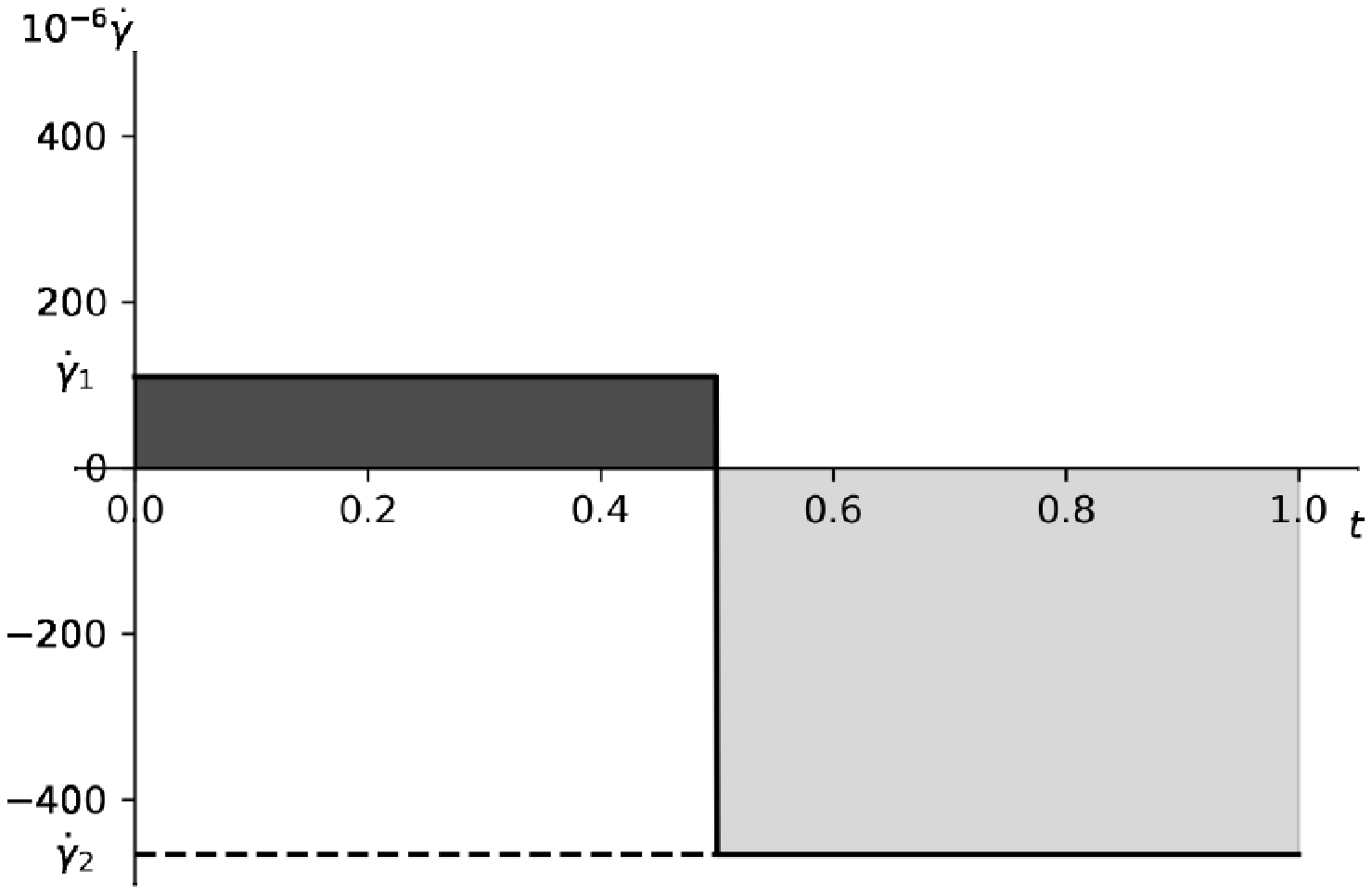}
(b)\includegraphics[width=0.47\columnwidth]{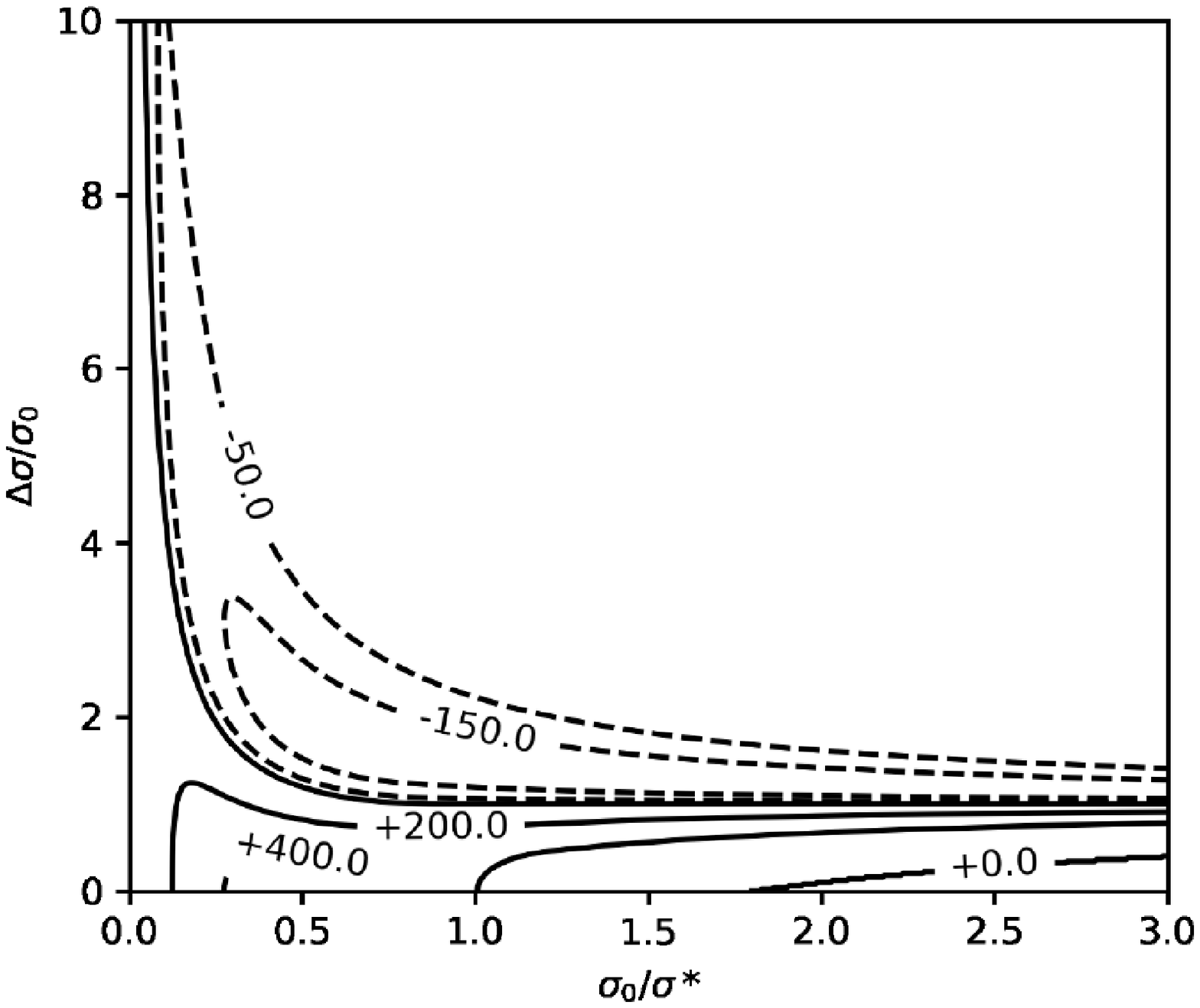}
\par\end{centering}
\caption{\label{fig:(a)square_wave_stress_phi0.6}(a) Strain rate $\dot{\gamma}$
evolution of $\phi=0.6$ over a period of square-wave stress with
$\sigma_{0}/\sigma^{*}=0.35,\Delta\sigma/\sigma_{0}=2.4$. (b) Average
shear rates $10^{-6}\bar{\dot{\gamma}}$ of suspension of volume fraction
$\phi=0.6$ under square wave stresses of different average stresses
$\sigma_{0}/\sigma^{*}$ and oscillation amplitudes $\Delta\sigma/\sigma_{0}$.}
\end{figure}

From Eq. \ref{eq:square_stress}, we see that shear stress alternates
between $\sigma_{1}=\sigma_{0}+\Delta\sigma$ and $\sigma_{2}=\sigma_{0}-\Delta\sigma$
over equal time periods. The shear rate evolution over a period is
shown in Fig. \ref{fig:(a)square_wave_stress_phi0.6}(a).

Fig. \ref{fig:(a)square_wave_stress_phi0.6}(b) shows average shear
rates $\bar{\dot{\gamma}}$ under different $\sigma_{0},\Delta\sigma$.
Negative shear rates can only be achieved for $\Delta\sigma/\sigma_{0}>1$
as this is the minimum condition for any negative stress to appear
at all. For higher $\Delta\sigma/\sigma_{0}$ and $\sigma_{0}/\sigma^{*}$
however, there can be zero shear rate where the system is jammed.

\begin{figure}[H]
(a)%
\begin{minipage}[t]{0.49\columnwidth}%
\includegraphics[scale=0.47]{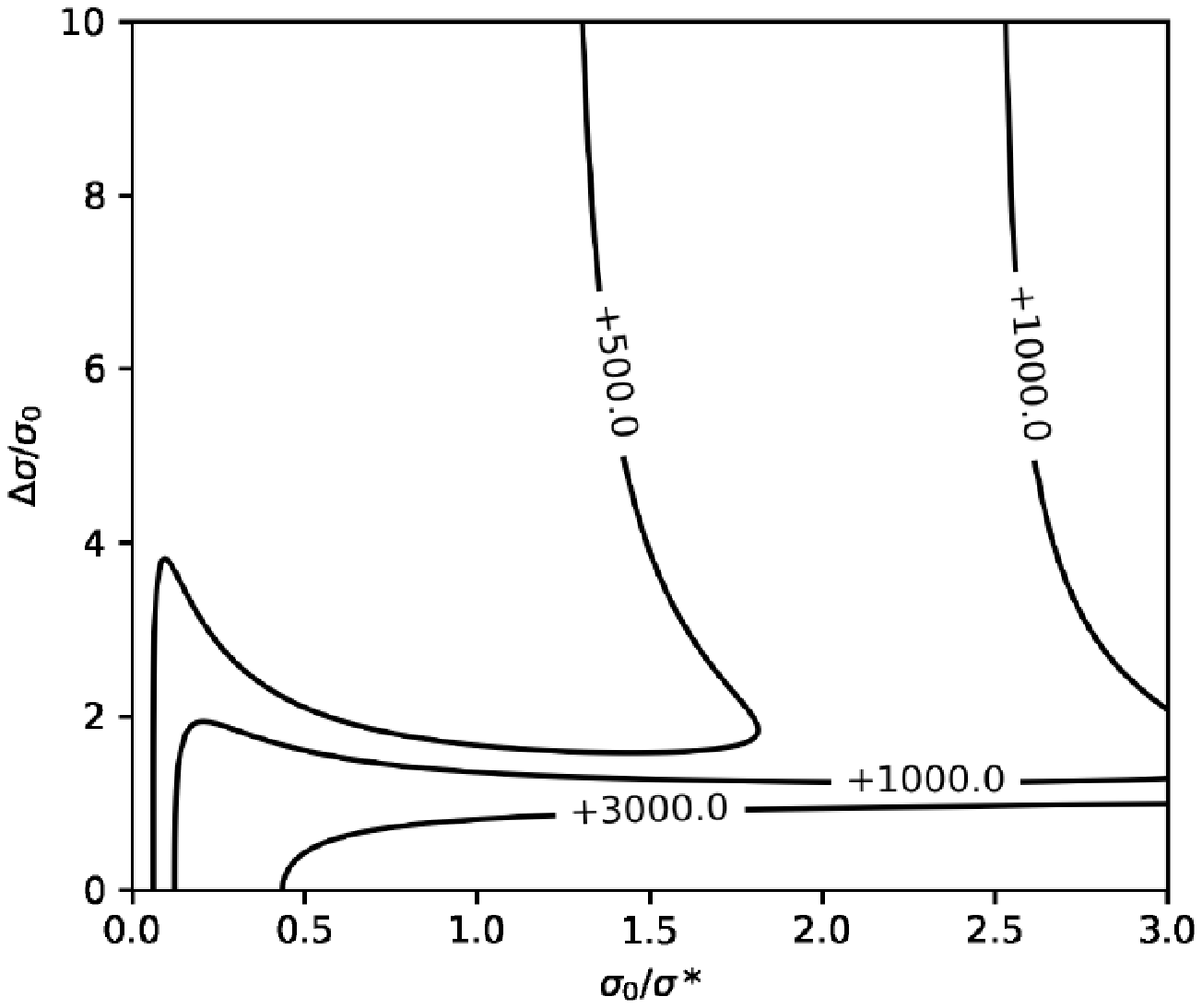}%
\end{minipage}\hfill{}(b)%
\begin{minipage}[t]{0.49\columnwidth}%
\includegraphics[scale=0.47]{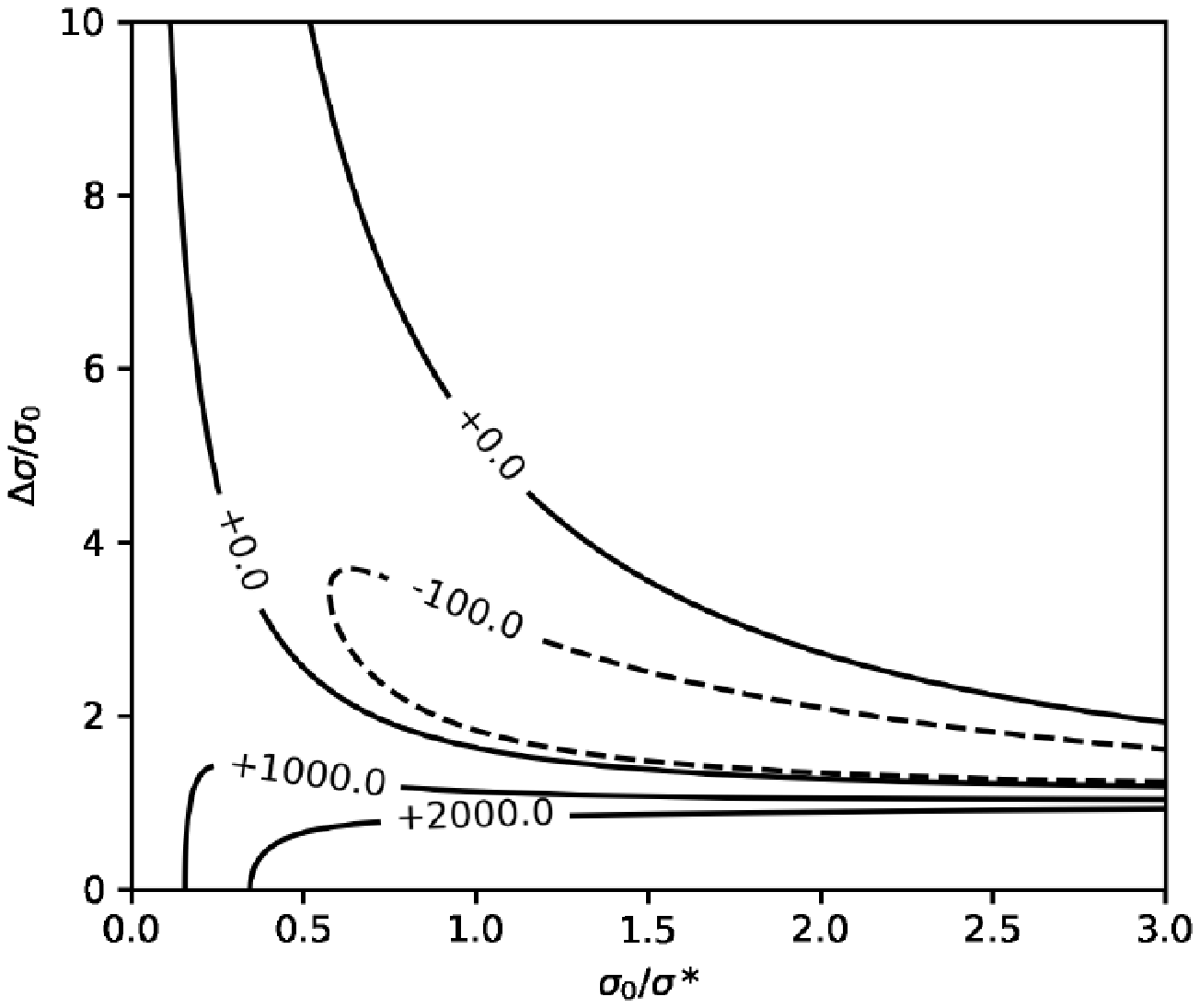}%
\end{minipage}

(c)%
\begin{minipage}[t]{0.49\columnwidth}%
\includegraphics[scale=0.47]{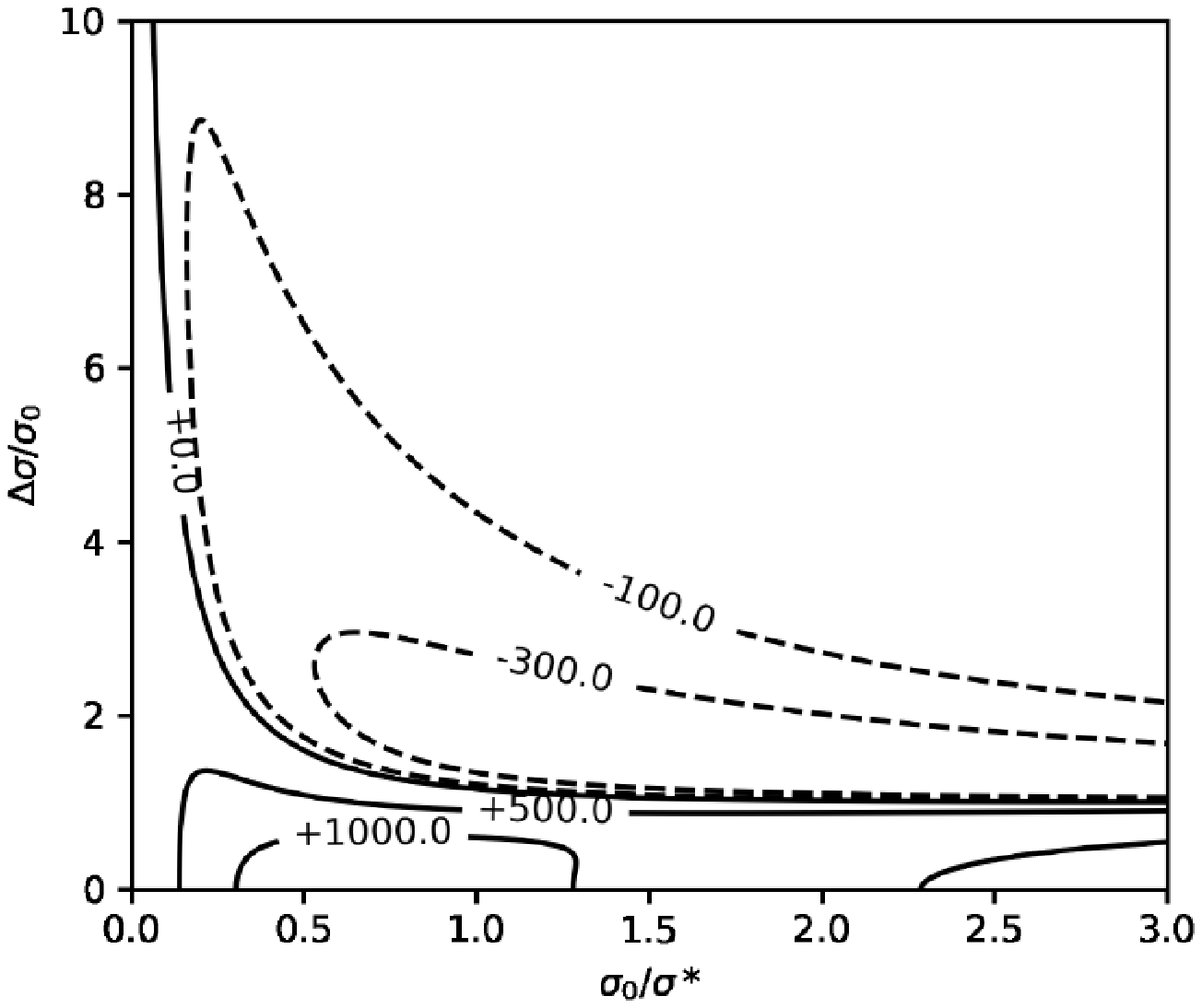}%
\end{minipage}\hfill{}(d)%
\begin{minipage}[t]{0.49\columnwidth}%
\includegraphics[scale=0.47]{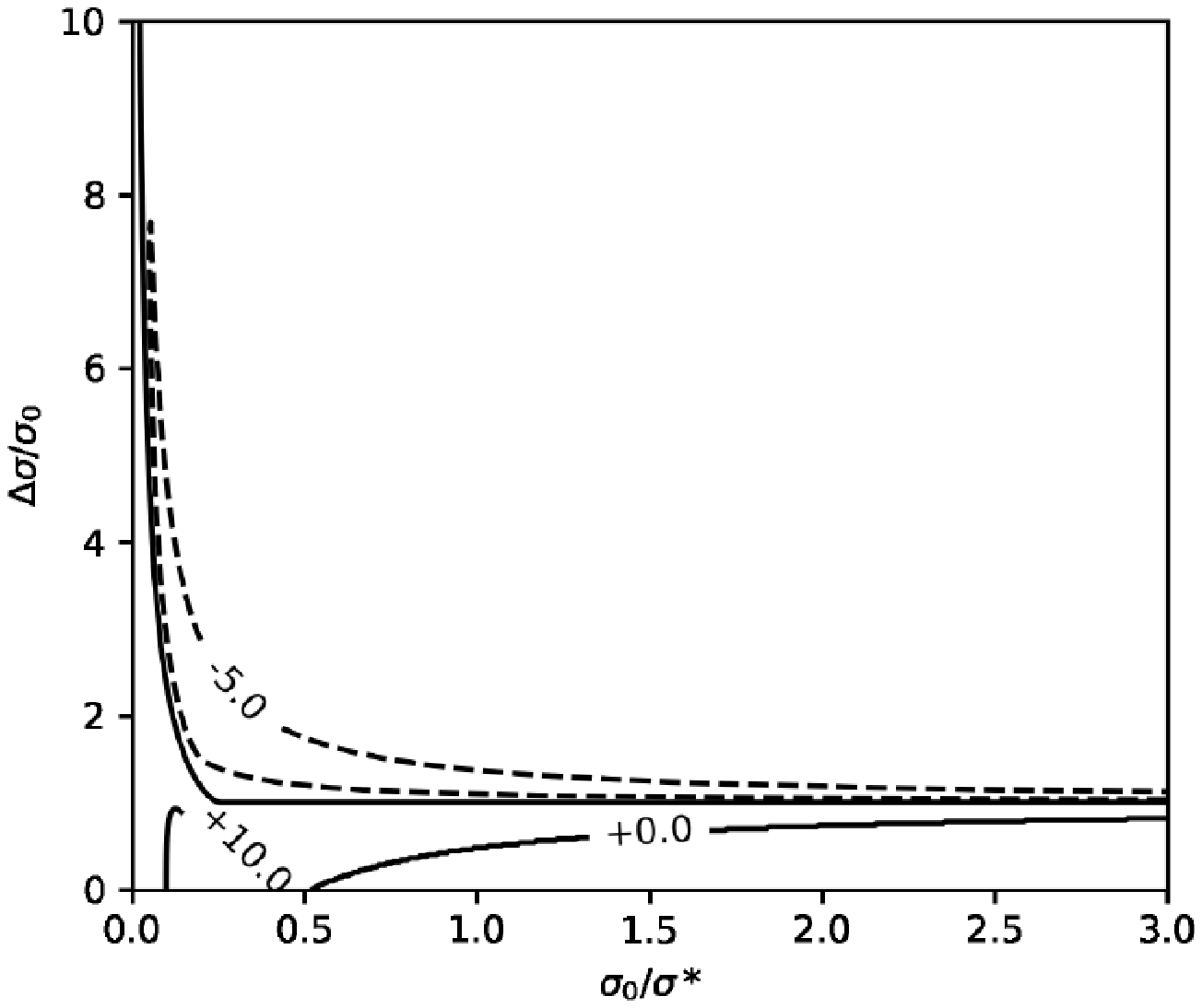}%
\end{minipage}

\caption{\label{fig:Square stress with different phi}Average shear rates $10^{-6}\bar{\dot{\gamma}}$
of suspensions of volume fractions $\phi$: (a) 0.55 (b) 0.56 (c)
0.58 (d) 0.63 under square wave stresses of different average stresses
$\sigma_{0}/\sigma^{*}$ and oscillation amplitudes $\Delta\sigma/\sigma_{0}$.}
\end{figure}

As shown in Fig. \ref{fig:Square stress with different phi}, negative
shear rates cannot occur for $\phi=0.55<0.555$ without DST, while
they can occur for $0.555<\phi<0.64$ with DST or SJ. One key distinction
between the DST, and SJ regime is that for DST materials the band
of conditions where negative shear rates are seen is bounded by regions
where the average shear rate is always positive (c.f. Fig. \ref{fig:Square stress with different phi}(b)),
where SJ fluids lack the second space of positive shear rates (c.f.
Fig. \ref{fig:Square stress with different phi}(c)-(d)). For DST
fluids the transition back to positive shear rates corresponds to
scenarios where positive shear rates along the high friction branch
dominate, and this branch of the flow curve does not exist where shear
jamming takes place.

In Fig. \ref{fig:Square stress with different phi}(c), we see the
transition from DST to SJ, for which the combination of large stresses
and large oscillations always produces a negative shear rate with
absolute value tending towards zero. Simply put, when the stresses
become large, the fluid will be very nearly jammed under both positive
and negative stresses, but on balance it will still be more flowable
for negative stresses.

For SJ systems $0.57<\phi<0.64$ (c.f. Fig. \ref{fig:Square stress with different phi}(d))
sufficiently large stresses are capable of inducing jamming, i.e.
the total cessation of flow $\dot{\gamma}=0$. The bounding contour
for total jamming, $\bar{\dot{\gamma}}=0$, is not shown.

\subsection{\label{subsec:Sine-wave-stress}Sine-wave Oscillations in Shear Stress}

Square wave stress protocols are useful for introducing the idea of
shear rates opposite to shear stresses in DST fluids, and here we
show that the same mechanism extends (albiet less efficiently) to
a flow driven by a sinusoidal stress protocol, $\sigma(t)=\sigma_{0}+\Delta\sigma\textup{sin}(2\pi t)$.
The stresses and shear rates at each moment are determined and the
shear rates are averaged with Eq. \ref{eq:averaging shear rates}.

\begin{figure}[H]
\begin{centering}
(a)%
\begin{minipage}[t]{0.49\columnwidth}%
\includegraphics[scale=0.47]{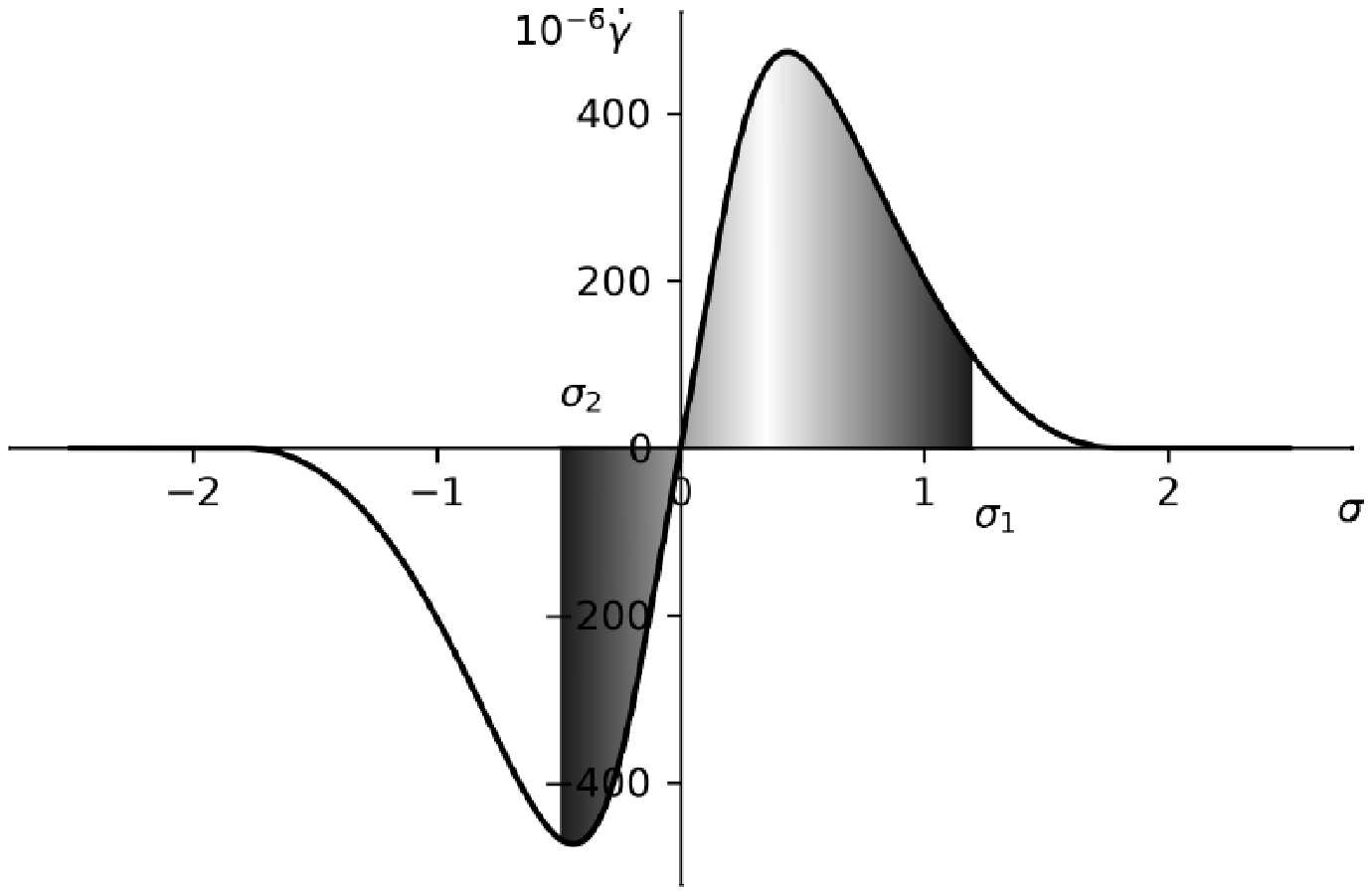}%
\end{minipage}(b)%
\begin{minipage}[t]{0.49\columnwidth}%
\includegraphics[scale=0.47]{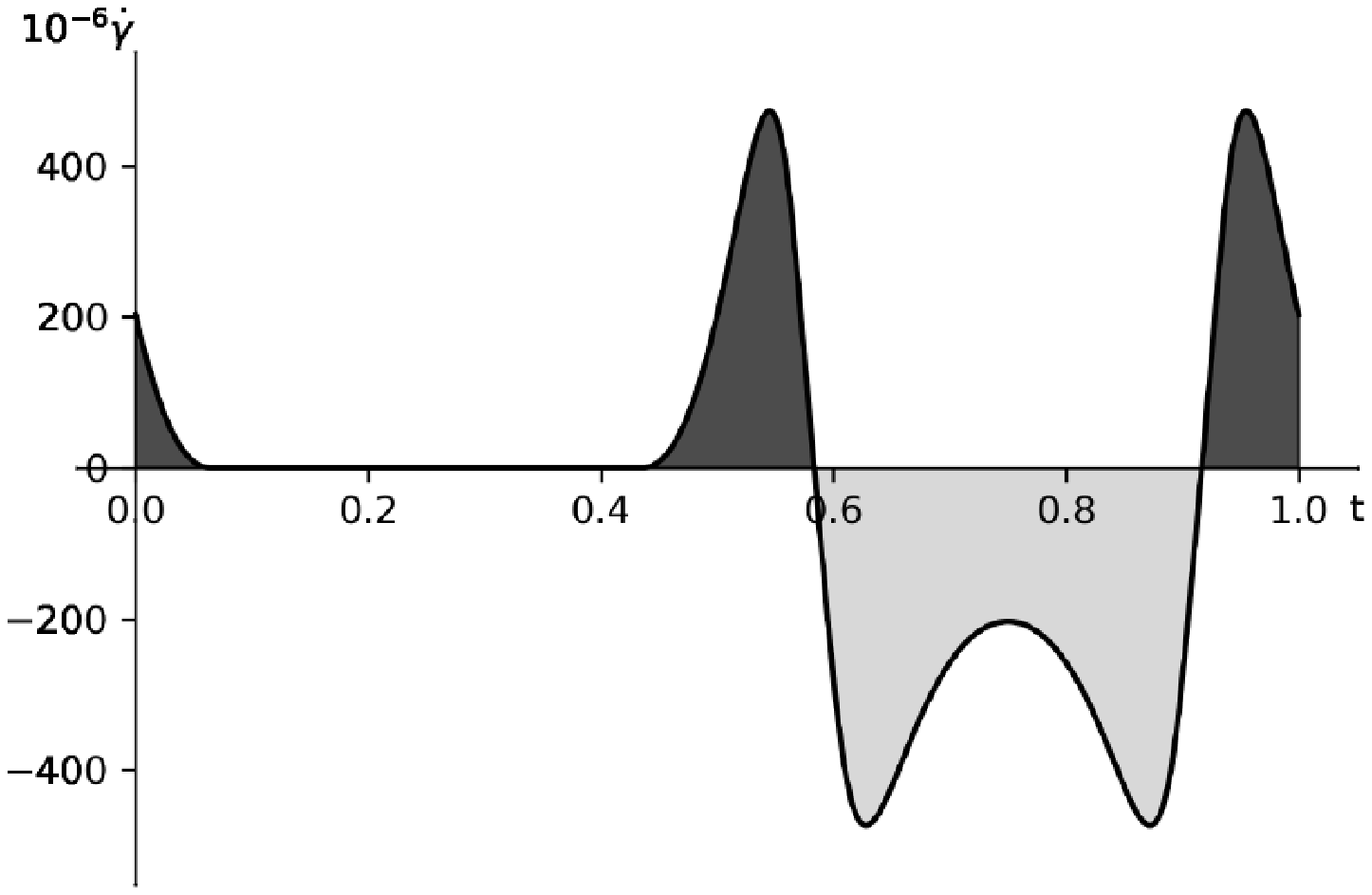}%
\end{minipage}
\par\end{centering}
\begin{centering}
(c)%
\noindent\begin{minipage}[t]{1\columnwidth}%
\begin{center}
\includegraphics[scale=0.47]{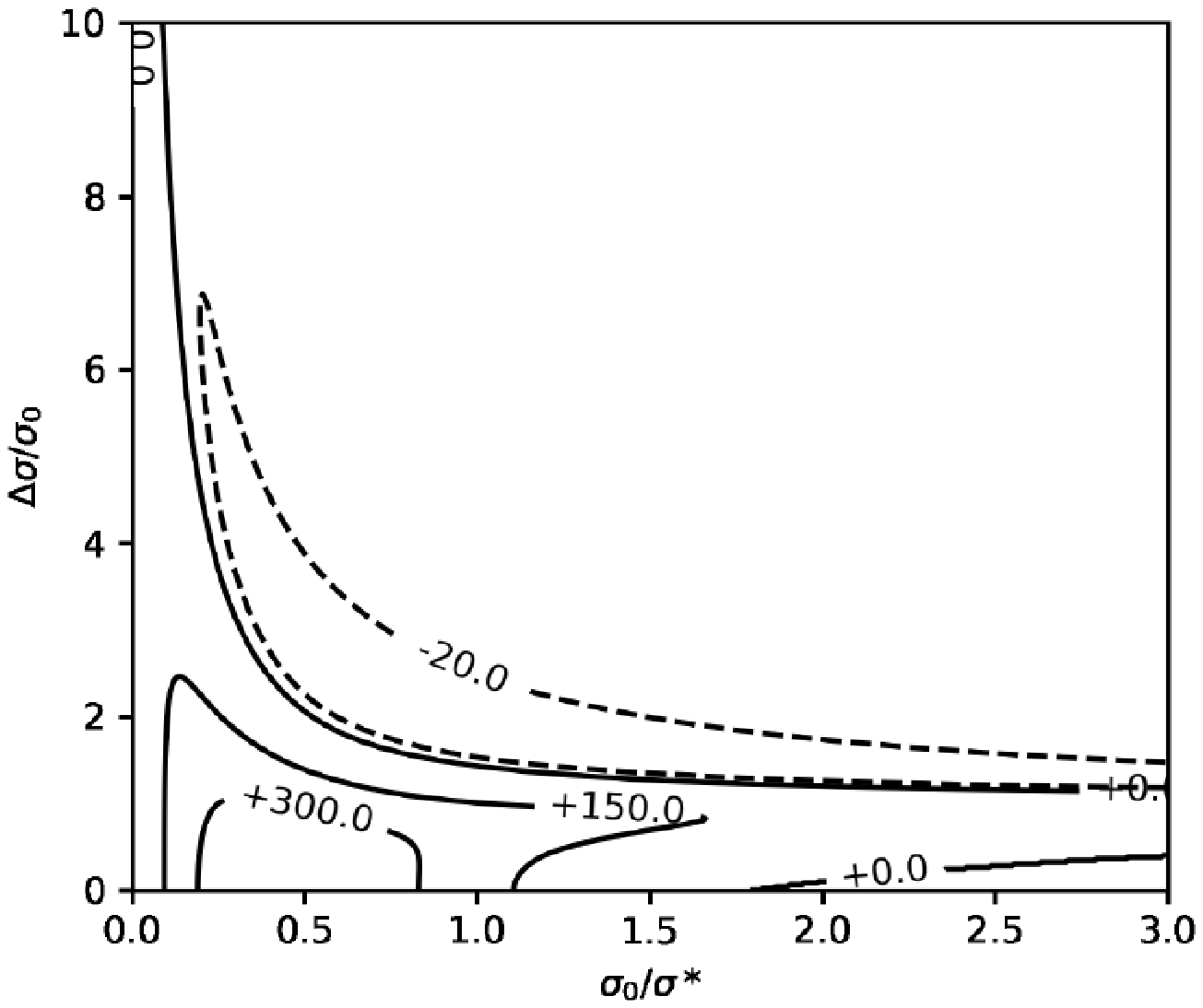}
\par\end{center}%
\end{minipage}
\par\end{centering}
\noindent \caption{\label{fig:Sinusoidal stress with flow curve}(a) The strain rate
$\dot{\gamma}$ - stress $\sigma$ relationship for suspension with
$\phi=0.6$. The colour gradient demonstrates how much different stresses
are weighted when averaging strain rates over a period. (b) The strain
rate $\dot{\gamma}$ evolution over a period of sinusoidal stress
with average stress $\sigma_{0}/\sigma^{*}=1$, relative amplitude
$\Delta\sigma/\sigma_{0}=2$. (c) Average shear rates $10^{-6}\bar{\dot{\gamma}}$
of suspension of volume fraction $\phi=0.6$ under sinusoidal stresses
of different average stresses $\sigma_{0}/\sigma^{*}$ and relative
amplitude $\Delta\sigma/\sigma_{0}$.}
\end{figure}

As shown in Fig. \ref{fig:Sinusoidal stress with flow curve}, the
sine wave stress protocol gives predictions similar to those observed
in Fig. \ref{fig:(a)square_wave_stress_phi0.6}. Here, however, it
is harder for negative average shear rates to occur because the fluid
spends more time exploring the non-inverted portions of the flow curve
with high positive shear rates.

\begin{figure}[H]
(a)%
\begin{minipage}[t]{0.49\columnwidth}%
\includegraphics[scale=0.47]{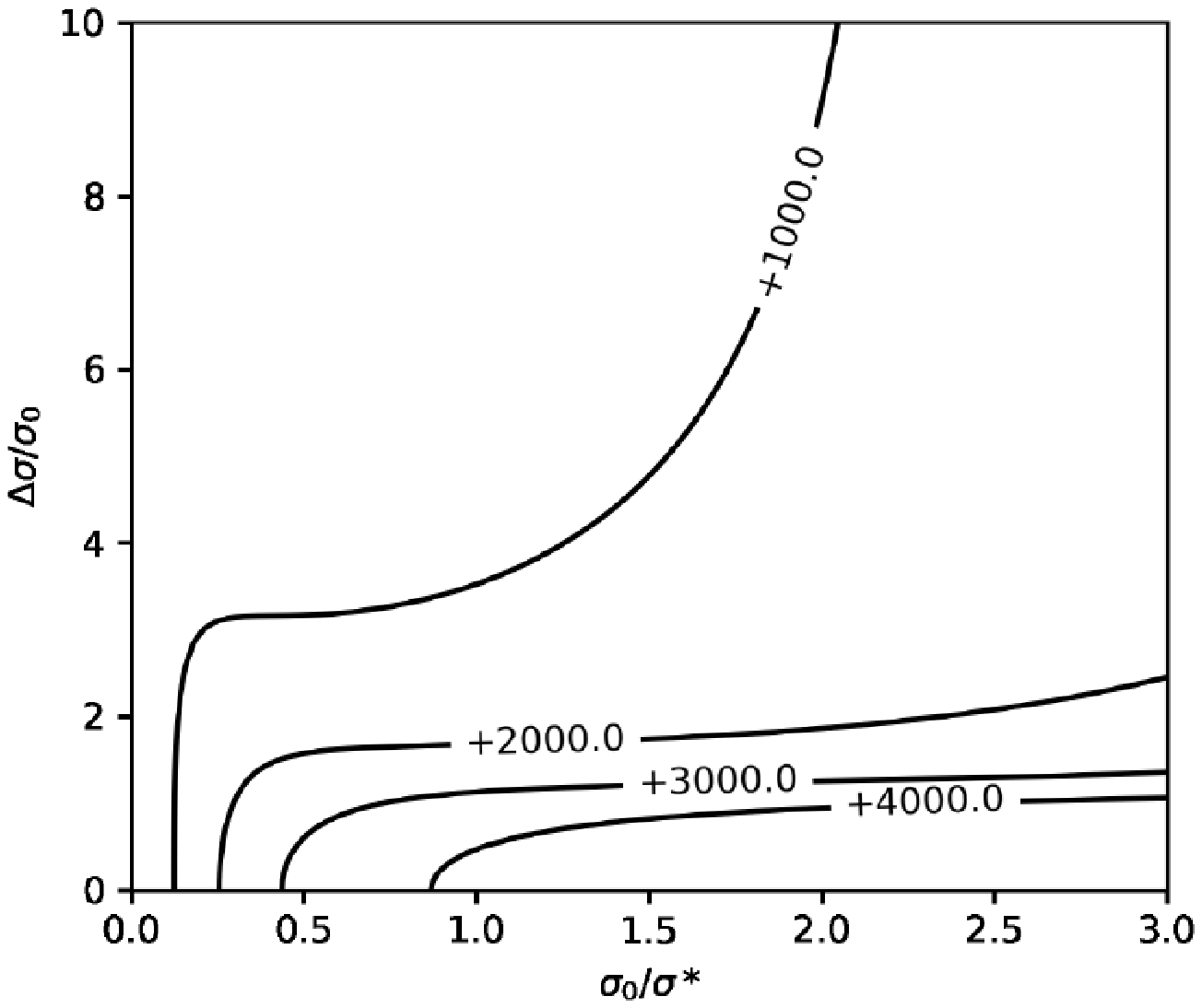}%
\end{minipage}\hfill{}(b)%
\begin{minipage}[t]{0.49\columnwidth}%
\includegraphics[scale=0.47]{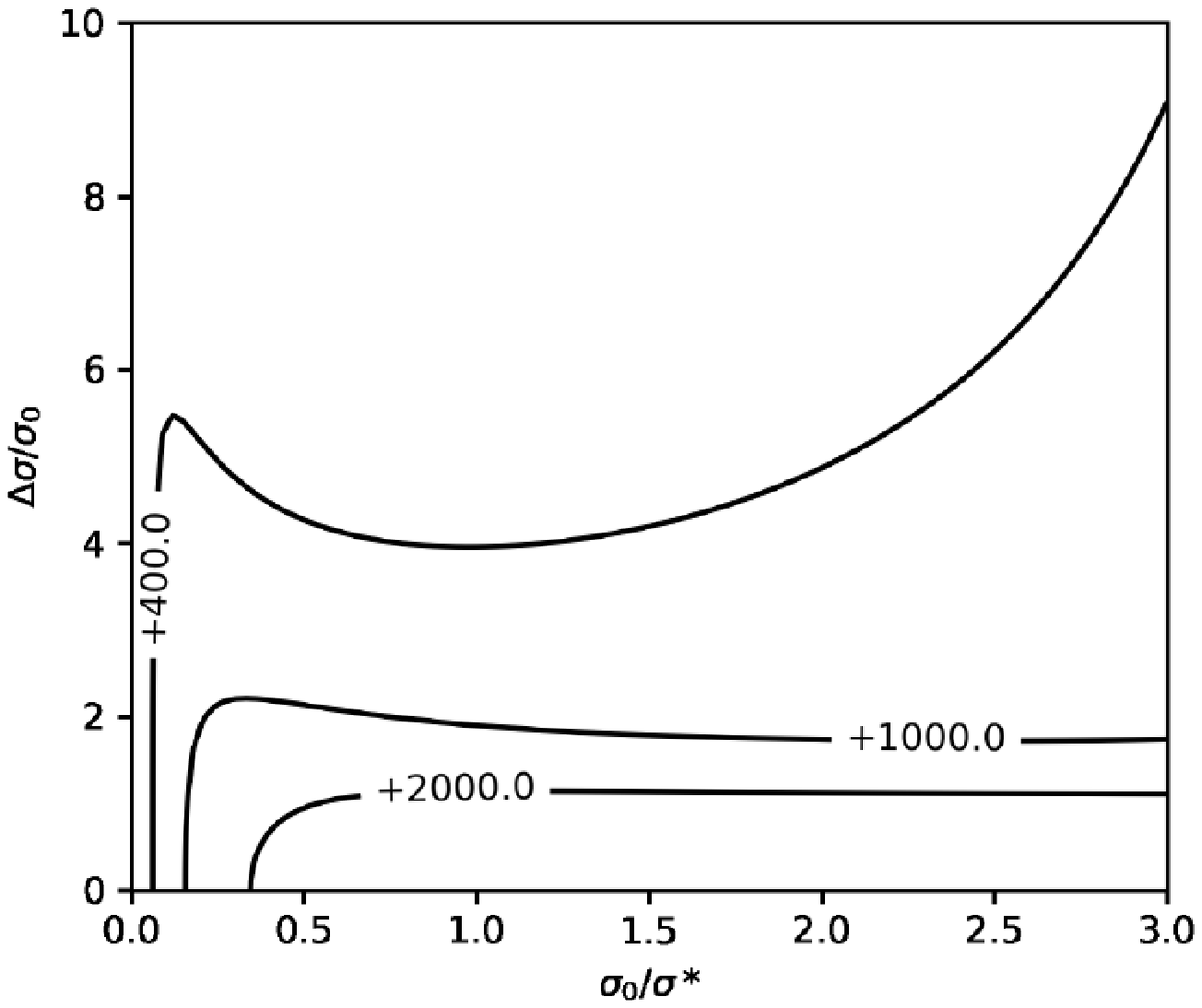}%
\end{minipage}

(c)%
\begin{minipage}[t]{0.49\columnwidth}%
\includegraphics[scale=0.47]{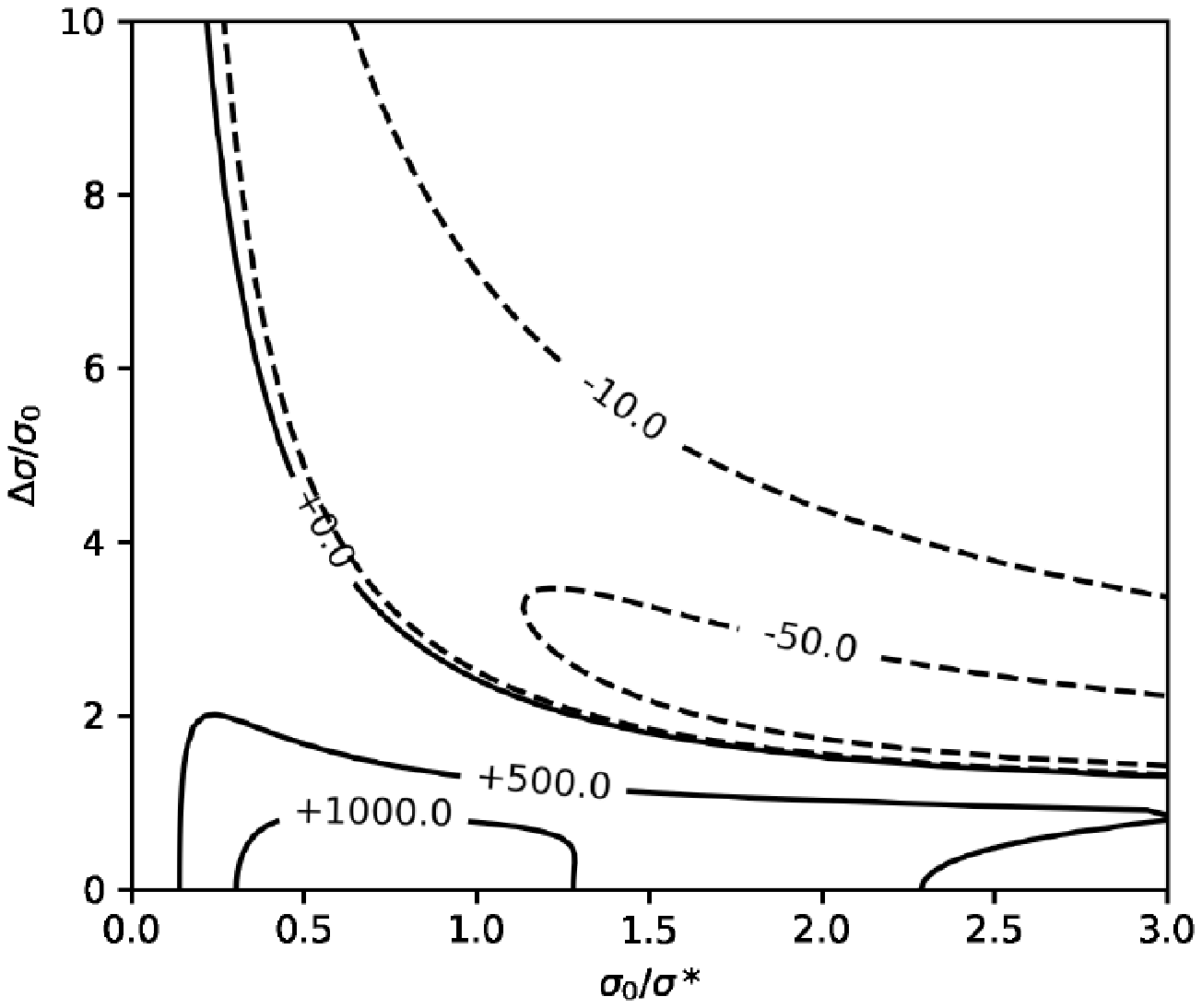}%
\end{minipage}\hfill{}(d)%
\begin{minipage}[t]{0.49\columnwidth}%
\includegraphics[scale=0.47]{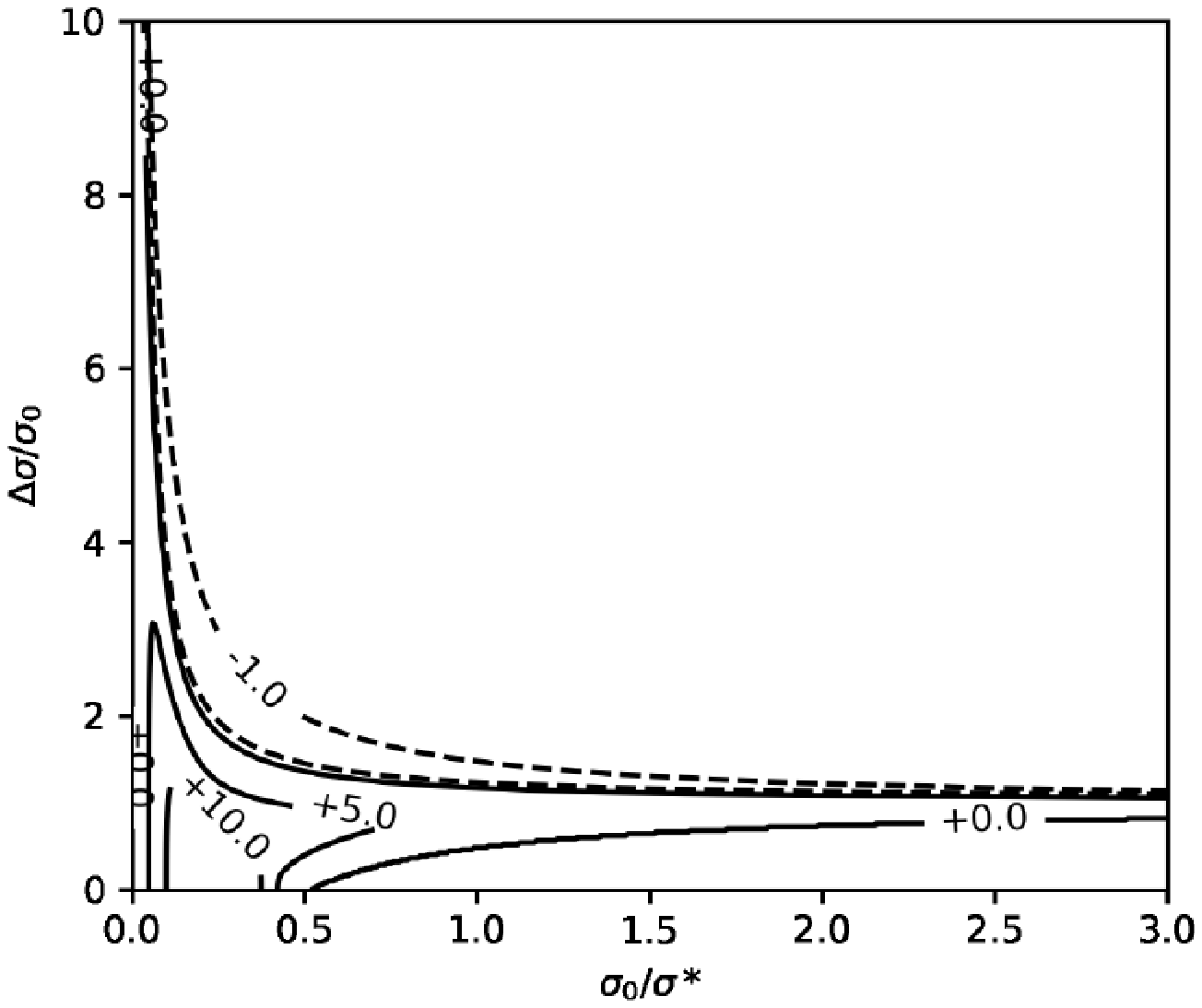}%
\end{minipage}

\caption{\label{fig:Sinusoidal stress with different phi}Average shear rates
$\dot{\gamma}$ of suspensions of volume fractions $\phi$: (a) 0.55
(b) 0.56 (c) 0.58 (d) 0.63 under sinusoidal stresses of different
average stresses $\sigma_{0}/\sigma^{*}$ and relative amplitudes
$\Delta\sigma/\sigma_{0}$.}
\end{figure}

As shown in Fig. \ref{fig:Sinusoidal stress with different phi},
negative shear rates can only be observed for $\phi>0.57$ when the
stress protocol is sinusoidal. This is a stricter requirement than
the square wave, which only requires DST or $\phi>0.555$. At high
$\phi$, the inverted portion of the flow curve is more extended,
so large positive shear rates will be explored less by the fluid and
it is easier to achieve negative average shear rates.

\subsection{\label{subsec:square_gravity}Square-wave Oscillations in Gravity}

Simple shear flows driven by defined stresses are perhaps the most
well-defined means of testing the premise of a ``negative viscosity''
in vibrated dense suspensions. In this subsection (and the one that
follows) we turn our attention to predictions for falling films in
oscillatory gravity. The basic premise is still the same: the strength
of gravity controls the scale of stresses within the fluid, and there
are conditions where the fluid moves faster under weak (reversed)
gravity than strong (normal) gravity.

In parallel to our analysis in section \ref{subsec:Square-wave-stress},
we first consider the motion of the fluid film under a gravity following
a square-wave variation in time:

\begin{equation}
g_{x}(t)=\begin{cases}
g_{0}+\Delta g & \text{if }t\text{ (mod }1)<1/2\\
g_{0}-\Delta g & \text{if }t\text{ (mod }1)\geq1/2
\end{cases}\label{eq:square_gravity}
\end{equation}

The motion of the fluid film is described by Eq. \ref{eq:NS_1D_ndim}.
We first assume Re $\sim0$, so:

\begin{equation}
1\pm\frac{\Delta g}{g_{0}}=\frac{\partial\sigma}{\partial x}\label{eq:NS_ND_square gravity}
\end{equation}

Taking the free surface as $x=0$, we have:

\begin{equation}
\sigma=(1\pm\frac{\Delta g}{g_{0}})x\label{eq:stress distribution under square gravity}
\end{equation}

which gives the shear stress at every position of the film $x\in[0,1]$.
The shear rates at each $x$ position are determined from the corresponding
shear stresses. The velocity is obtained through integrating shear
rates:

\begin{equation}
v(x)=\int_{0}^{x}dx'\dot{\gamma}(t,x)+u(t)\label{eq:intergrating shear rates}
\end{equation}

where $u(t)$ is the velocity of the solid boundary, and over a period
$\bar{u(t)}=0$. Averaging $v(x)$ over each position $x\in[0,1]$,
we can obtain the velocity of the film:

\begin{equation}
\bar{u}=\int_{0}^{1}dxu(t,x)\label{eq:Averaing velocity in fluid film}
\end{equation}

\begin{figure}
(a)%
\begin{minipage}[t]{0.49\columnwidth}%
\includegraphics[scale=0.47]{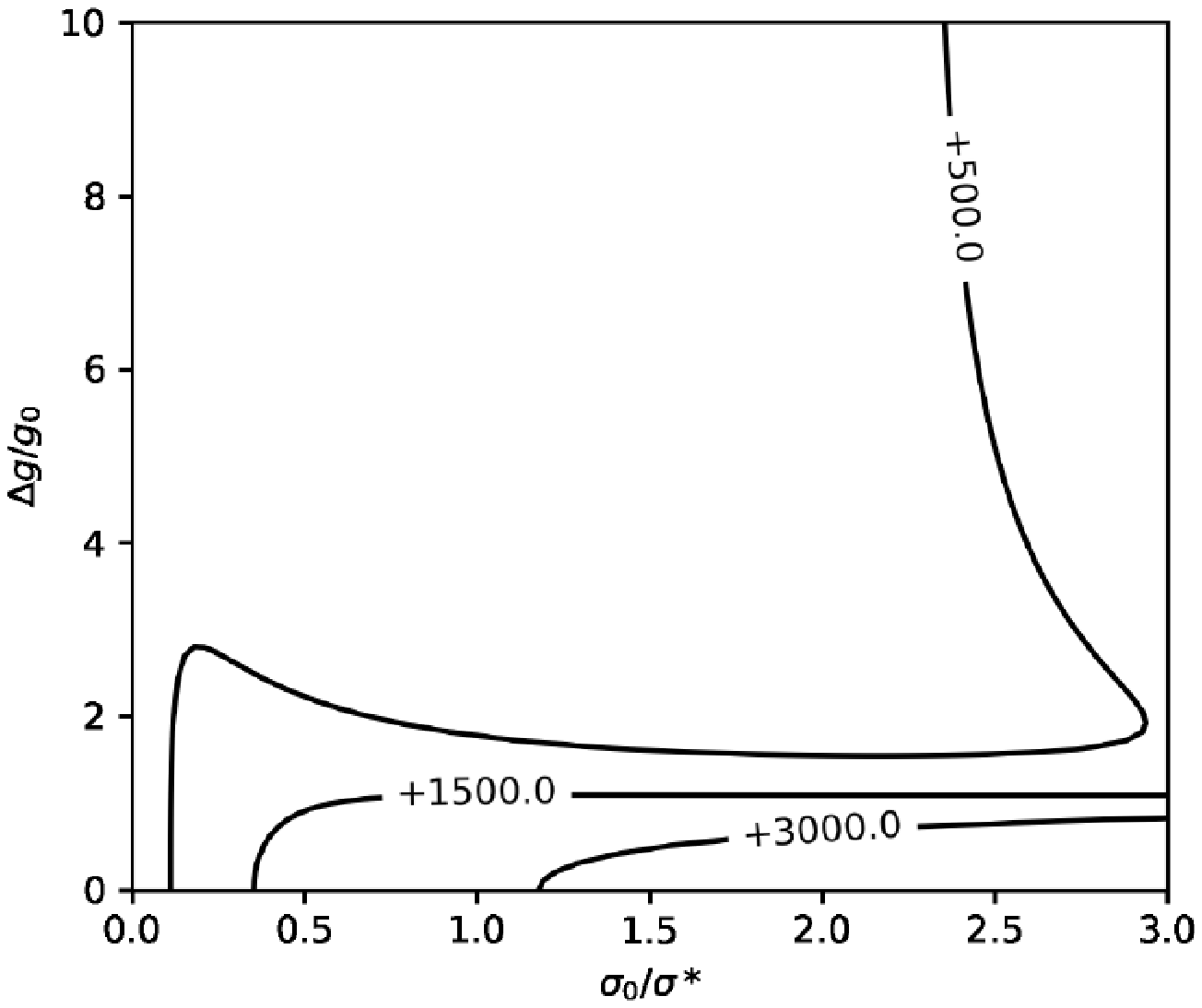}%
\end{minipage}\hfill{}(b)%
\begin{minipage}[t]{0.49\columnwidth}%
\includegraphics[scale=0.47]{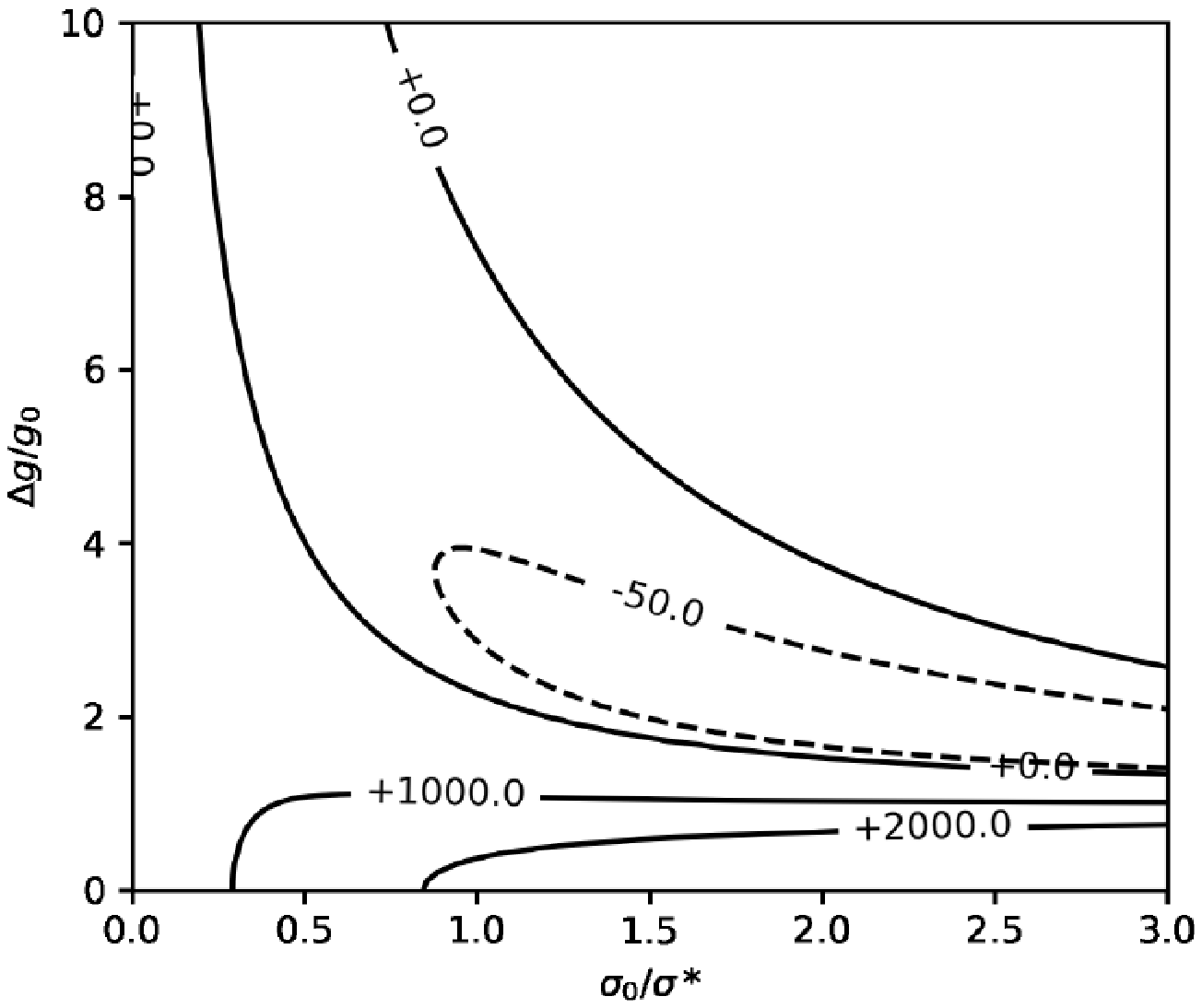}%
\end{minipage}

(c)%
\begin{minipage}[t]{0.49\columnwidth}%
\includegraphics[scale=0.47]{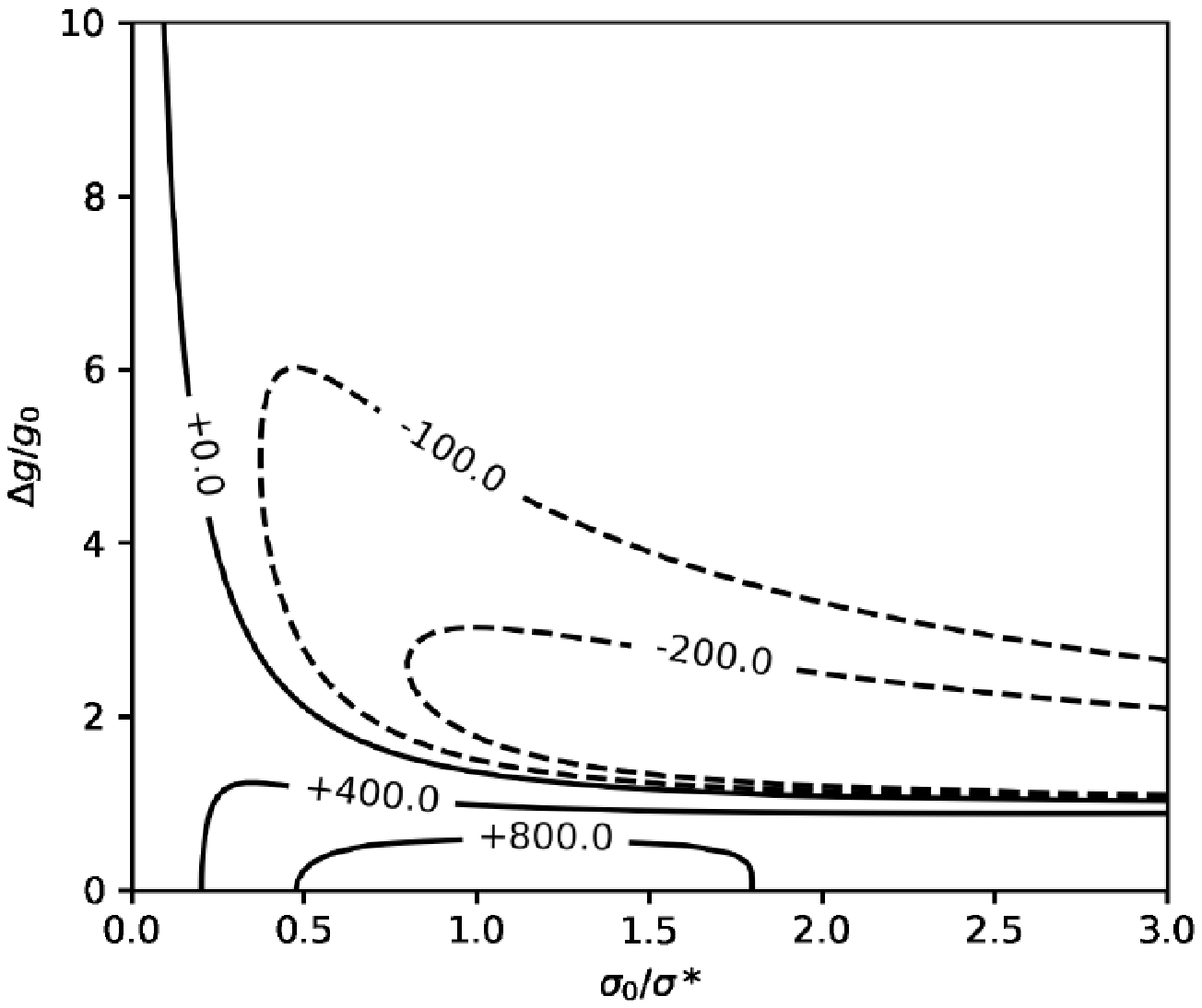}%
\end{minipage}\hfill{}(d)%
\begin{minipage}[t]{0.49\columnwidth}%
\includegraphics[scale=0.47]{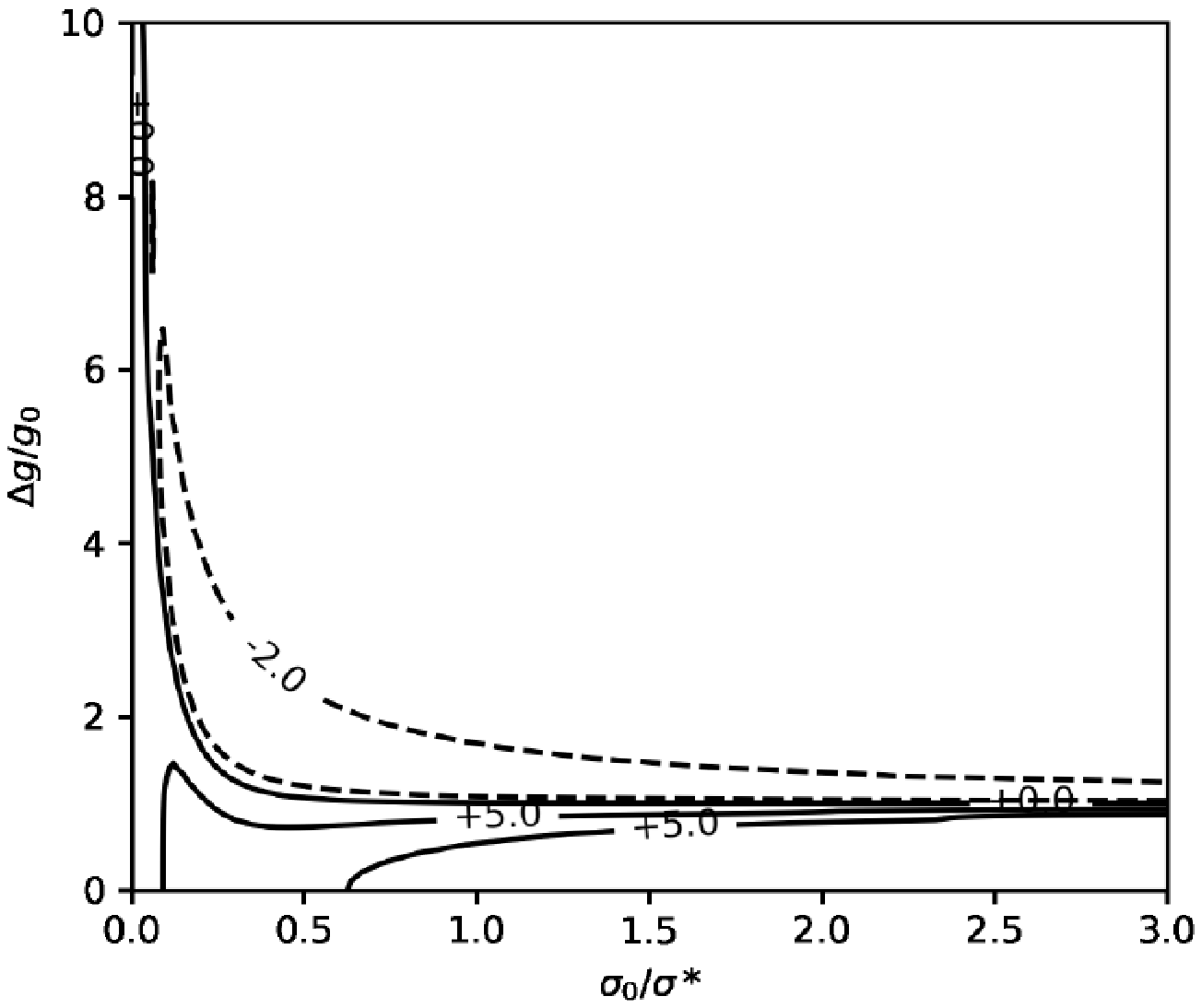}%
\end{minipage}

\caption{\label{fig:Square droplet}The velocities $10^{-6}\bar{u}$ of falling
films of volume fractions $\phi$: (a) 0.55 (b) 0.56 (c) 0.58 (d)
0.63 under square-wave gravity of different average stresses $\text{\ensuremath{\sigma_{0}}/\ensuremath{\sigma^{*}}}$
(where $\sigma_{0}=\rho g_{0}H$) and relative amplitudes $\Delta g/g_{0}$.}
\end{figure}

The average velocity over a period is obtained through Eq. \ref{eq:averaging shear rates}.
Similar to section \ref{subsec:square_gravity}, negative values can
be obtained for $\Delta g/g_{0}>1$, which is the minimum condition
for any stress to appear at all. However, Eq. \ref{eq:stress distribution under square gravity}
shows that part of the fluid is constantly under low stress, i.e.,
in the uninverted portions of the flow curve. This makes it harder
to achieve negative $\bar{u}$ as the positive shear rates are explored
more by the fluid film.

\subsection{\label{subsec:sine_gravity}Sine-wave oscillations in Gravity}

In parallel with the results given in section \ref{subsec:Sine-wave-stress},
we now extend our analysis to fluid films with gravitational forcing
varying sinusoidally in time, $g(t)=g_{0}+\Delta g\textup{sin}(2\pi t)$.
In the case of sine waves, Eqs. \ref{eq:NS_ND_square gravity} and
\ref{eq:stress distribution under square gravity} are substituted
with

\begin{equation}
1\pm\frac{\Delta g}{g_{0}}\textup{cos}(t)=\frac{\partial\sigma}{\partial x},\sigma=[1\pm\frac{\Delta g}{g_{0}}\textup{cos}(t)]x\label{eq:fluid film sinusoidal stress distribution}
\end{equation}

The average velocity of the fluid film follows similar caclulation
procedure as in Eqs. \ref{eq:intergrating shear rates} and \ref{eq:Averaing velocity in fluid film}.
It is harder to achieve negative velocity than in both sections \ref{subsec:Sine-wave-stress}
and \ref{subsec:square_gravity}, as (1) similar to section \ref{subsec:square_gravity},
part of the film is constantly under low stress and (2) the gravity
can fluctuate to lower values during oscillation, so the fluid will
explore more the positive shear rates.

\begin{figure}[H]
(a)%
\begin{minipage}[t]{0.49\columnwidth}%
\includegraphics[scale=0.47]{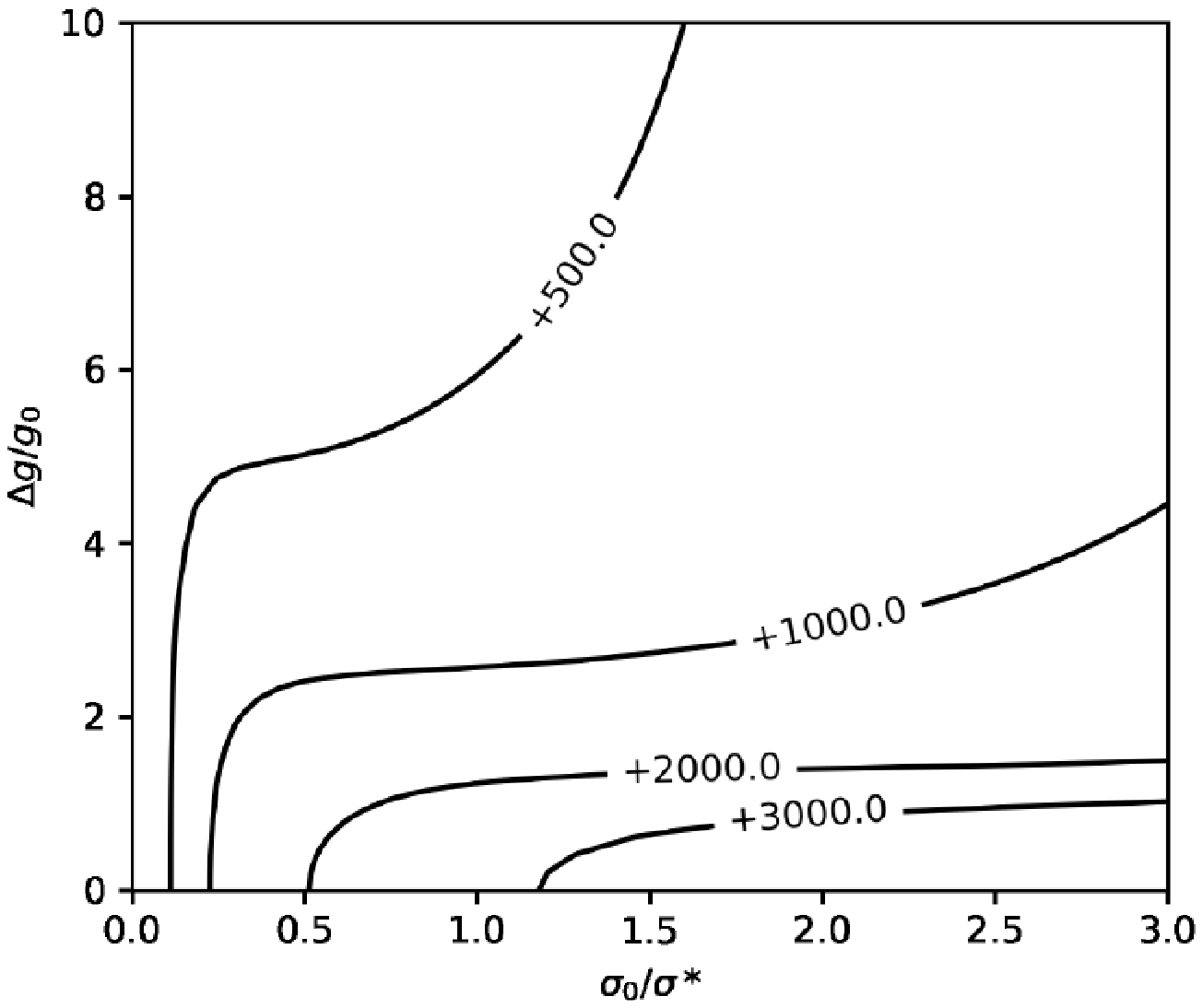}%
\end{minipage}\hfill{}(b)%
\begin{minipage}[t]{0.49\columnwidth}%
\includegraphics[scale=0.47]{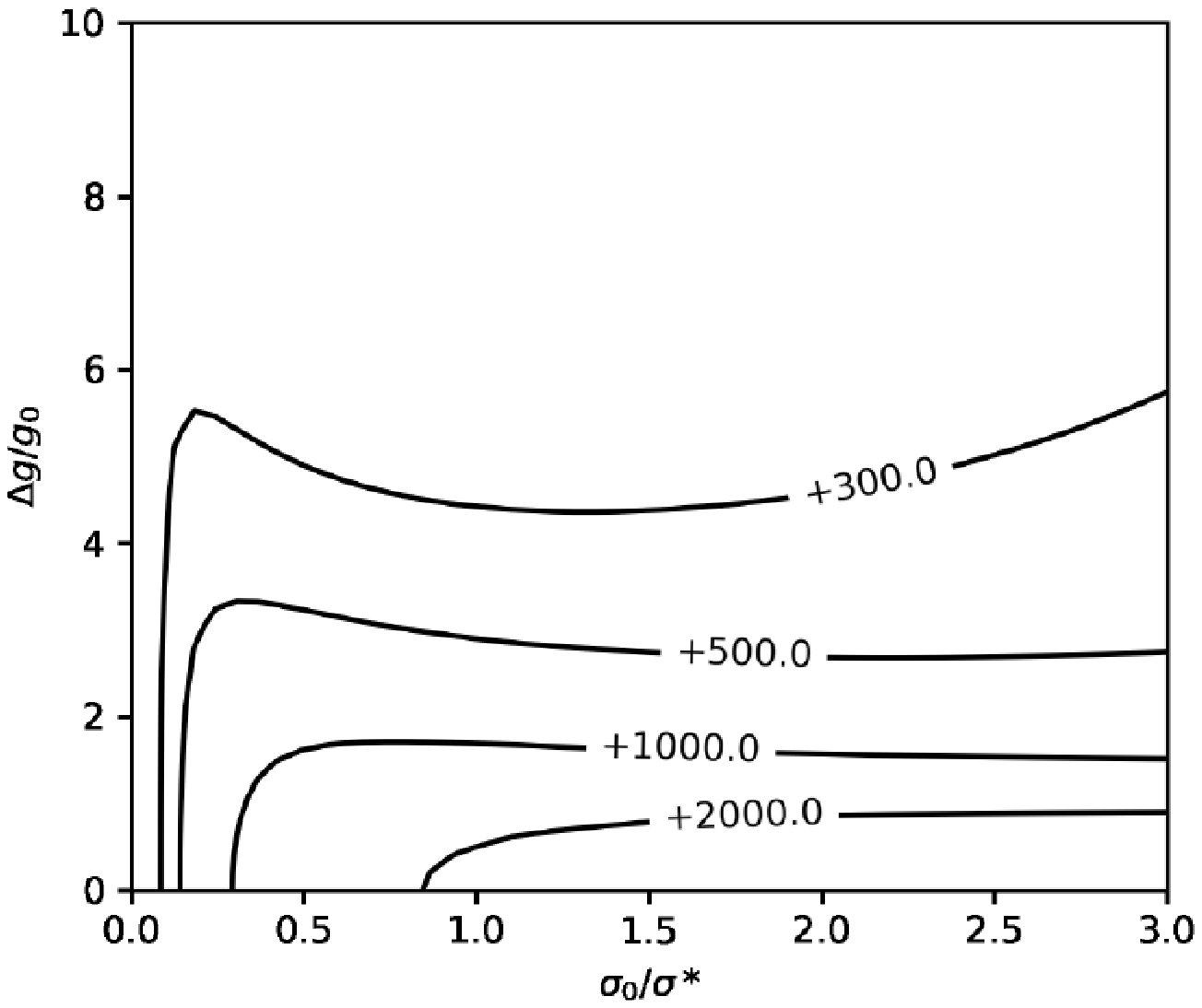}%
\end{minipage}

(c)%
\begin{minipage}[t]{0.49\columnwidth}%
\includegraphics[scale=0.47]{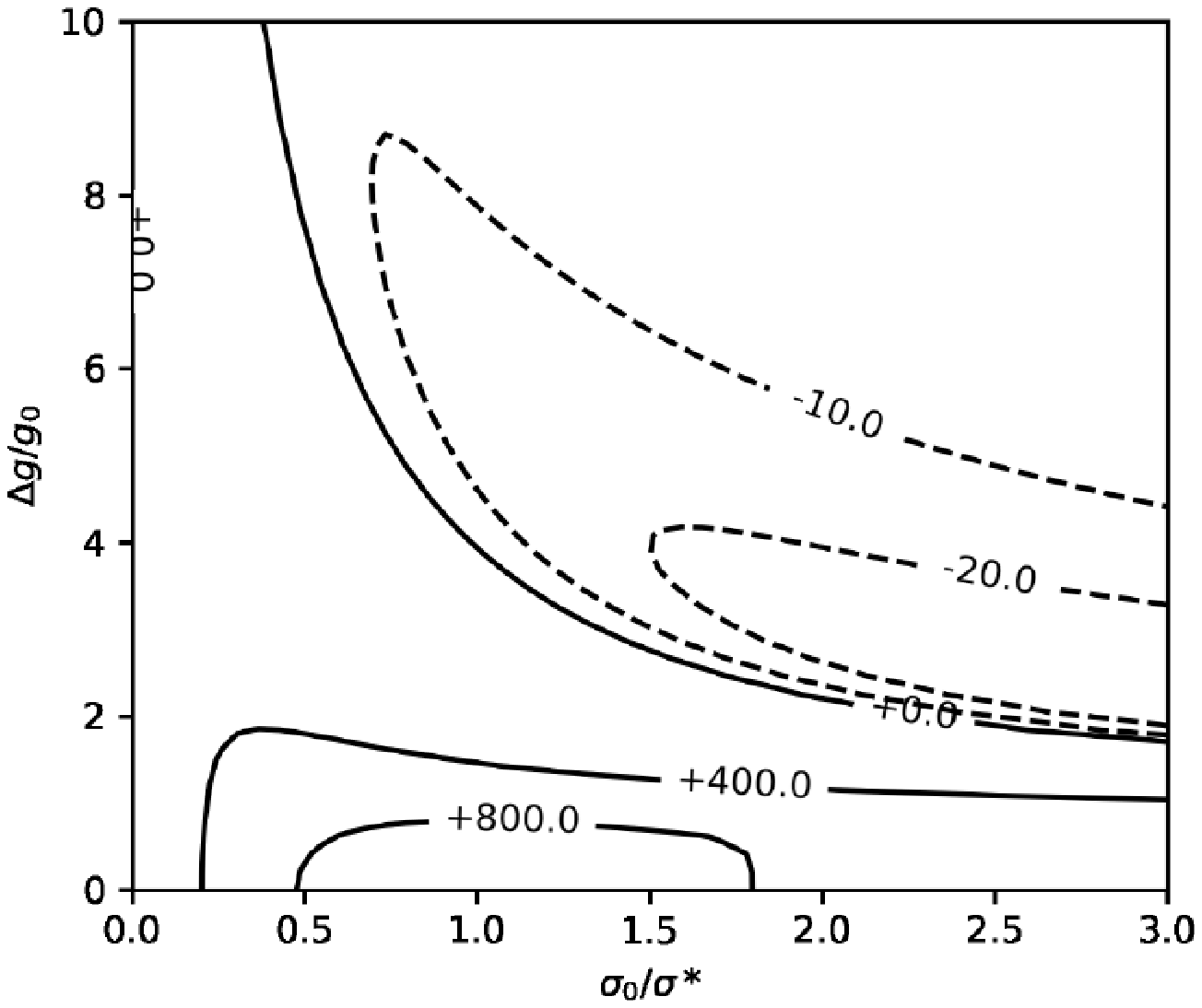}%
\end{minipage}\hfill{}(d)%
\begin{minipage}[t]{0.49\columnwidth}%
\includegraphics[scale=0.47]{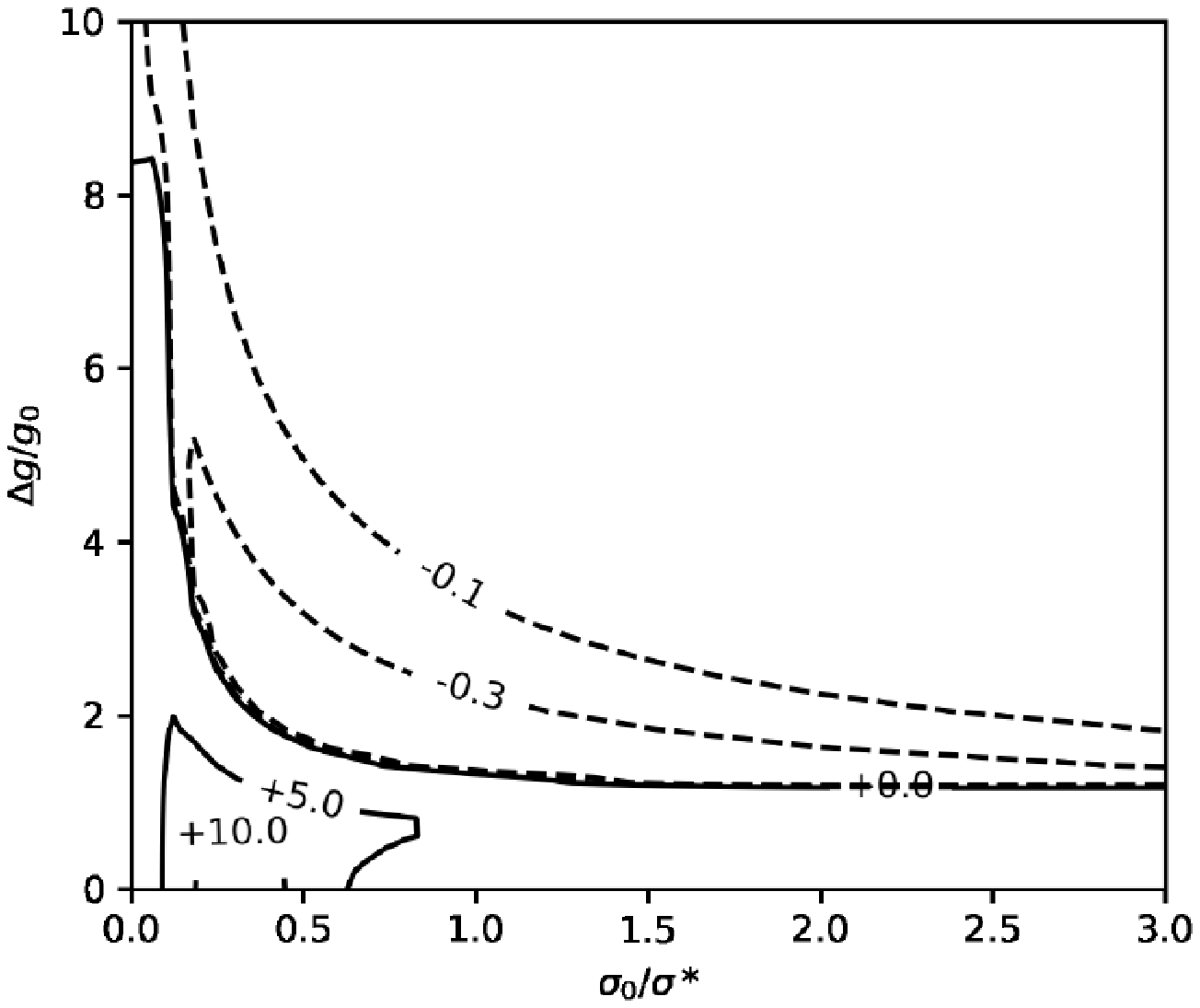}%
\end{minipage}

\caption{\label{fig:Sinusoidal droplet}Velocities $10^{-6}\bar{u}$ of films
of volume fractions $\phi$: (a) 0.55 (b) 0.56 (c) 0.58 (d) 0.63 under
sine-wave gravity of different average stresses $\text{\ensuremath{\sigma_{0}}/\ensuremath{\sigma^{*}}}$
(where $\sigma_{0}=\rho g_{0}H$) and relative amplitudes $\Delta g/g_{0}$.}
\end{figure}

\subsection{\label{subsec:intertia}Some results for finite inertia}

From Eqs. \ref{eq:NS_1D_ndim} and \ref{eq:Re expression}, the finite
Reynolds number can be of influence to the fluid motion. In the experiments,
$\eta_{\dot{\gamma}=0}\sim1Pa\cdot s$,$H\sim1mm$,$T\sim0.1s,\rho\sim1000kg/m^{3}$,
so Re $\sim0.01$. Here we consider an expansion for small Reynolds
numbers, Re$\ll1$, with sinusoidal variations in the graviational
field, as in section \ref{subsec:sine_gravity}. Our analysis will
not extend to consider the stability of a base-state limit cycle to
perturbations, where negative-sloping portions of the flow curve have
previously been shown to be unstable in steady shear \cite{mari2015nonmonotonic}.

Eq. (5) is now written as:

\begin{equation}
\text{Re}\ppt u=-(1+\frac{\Delta g}{g_{0}}\cos(2\pi t))+\frac{\partial\sigma}{\partial x}
\end{equation}

Expanding $u(t),\sigma(t)$ with respect to Re:

\begin{equation}
\begin{aligned}u(\text{Re},t)=u_{0}(t)+u_{1}(t)\text{Re}+...,\\
\sigma(\text{Re},t)=\sigma_{0}(t)+\sigma_{1}(t)\text{Re}+...
\end{aligned}
\end{equation}

Plugging (13) into (12) we find:

\begin{equation}
\begin{aligned}0= & \left\{ -\left[1+\frac{\Delta g}{g_{0}}\cos(2\pi t)\right]+\frac{\partial\sigma_{0}}{\partial x}\right\} \text{Re}^{0}+\left\{ -\frac{\partial u_{0}}{\partial t}+\frac{\partial\sigma_{1}}{\partial x}\right\} \text{Re}^{1}+...\end{aligned}
\end{equation}

The coefficients for each order of Re should be zero, so:

\begin{equation}
\frac{\partial\sigma_{0}}{\partial x}=1+\frac{\Delta g}{g_{0}}\cos(2\pi t),
\end{equation}

\begin{equation}
\frac{\partial\sigma_{1}}{\partial x}=\frac{\partial u_{0}}{\partial t}
\end{equation}

Eq. (15) is an even function about time $t=1/4$, the same for the
velocity $u_{0}(x,t)$ at any point $x$ under low Re. This also means
that $\partial_{t}u_{0}(x,t)$ is an odd function about time $t=1/4$
for any $x$. So it must also be true that $\sigma_{1}(x,t)=\int_{0}^{x}da\partial_{t}u_{0}(a,t)$
is an odd function about $t=1/4$, since it is just a sum of odd functions.

We also expand $\dot{\gamma}(t)$ with $\dot{\gamma}(Re,t)=\dot{\gamma_{0}}(t)+Re\dot{\gamma_{1}}(t)+...$.
For low Re we have:

\begin{equation}
\dot{\gamma}_{1}=\frac{\partial\dot{\gamma}}{\partial\sigma}(\sigma_{0}(x,t))\sigma_{1}
\end{equation}

Where the input to a function is even, then output must also be even,
so $\frac{\partial\dot{\gamma}}{\partial\sigma}(\sigma_{0}(t,y))$
is an even function about time $t=1/4$.

If we compute the average shear rate on a time $t\in[-1/4,3/4]$,
such that the averaging interval is centered about $t=1/4$, we are
taking the inner product of the odd $\sigma_{1}(x,t)$ function and
the even $\frac{\partial\dot{\gamma}}{\partial\sigma}(\sigma_{0}(t,y))$
function on the domain, so the integral evaluates to zero.

Thus, to a first correction, a small amount of inertia gives the first
correction term $\overline{\text{\ensuremath{\dot{\gamma}_{1}(t)}}}=0$.
Neglecting considerations of flow stability, the effect of inertia
should only appear in an $O(\text{Re}^{2})$ correction, with Re $\approx0.01$
in this case.

\section{\label{sec:Experimental-Results}Experimental Results}

In this section, we will first summarize what we observed in our experimental
study (section \ref{subsec:Summary-of-Experimental}) and then follow-up
with a discussion of those findings in relation to our experimental
observations (section \ref{subsec:Discussion-and-Relationship}).

\subsection{Summary of Experimental Observations\label{subsec:Summary-of-Experimental}}

The experimental setup has be explained in section \ref{subsec:Experiments}.
We used suspensions of cornstarch in water with varying particle mass
fractions between 52-55\%. To simulate the oscillating gravitational
field $\Delta g$, we modulate the frequency $f$ and amplitude $A$
of the oscillating speaker platform, $\Delta g=(2\pi f)^{2}A$, and
$g_{0}=9.8\text{m}/\text{s}^{2}$ is maintained as the constant value
of gravity. Our experiments allowed us to access frequencies as low
as $f=1$Hz and peak amplitudes as large as about $A=0.5$cm, but the simple
experimental setup did not permit direct control over the oscillation
amplitude $A$, and the power limations of our speaker/amplifier setup
meant that for frequencies above about $20$Hz the maximum displacement
amplitude decreased with increasing frequency. In the discusion that
follows, adjustments to the amplitude will be described qualitatively
in terms of speaker volume settings, in keeping with the level of
direct control we had over the experiments themselves.

Between $1\sim14$ Hz, the suspension develops Faraday waves when
the volume controls were set sufficiently high. For low frequency
$1\sim10$ Hz, Faraday waves are stable and the suspension does not
climb or form stable holes after being finitely perturbed. For frequencies
in the range of $12\sim14$ Hz, Faraday waves are unstable to finite-amplitude
perturbations, and the suspension can continuously climb up after
being perturbed.

Between $12\sim19$ Hz, columns of the suspension with thickness $\sim1$mm
and width $\sim3$mm can climb up vertical surfaces. At frequencies
greater than $20$ Hz, however, such climbing motion is slowed down,
possibly because our experimental setup lacks the power to maintain
large accelerations at high frequencies.

At $53\%$ mass fraction of cornstarch, droplets of thickness $1\sim2$
mm were observed to climb up at $2$mm/s under a sinusoidal signal
of 15Hz, 100\% volume, with an approximate $\Delta g\approx89m/s^{2}$,
corresponding to $\Delta g/g_{0}=8.9$. Droplets were formed on vertical
surfaces either without direct intervention (e.g. ``pinched off''
from climbing colums or splashing up from the reservoir below) or
by direct placement, collecting a small amount of fluid and allowing
it to wet the surface.

Holes and finger-like structures can appear in the range of $f=14\sim25$
Hz, as previously documented in earlier literature \cite{vonKann2014Phase}.
Within the lower end of this frequency range, $f=14\sim19$ Hz, the
delocalized holes and finger-like structures form quickly, rising
and falling periodically. At higher frequencies $20\sim25$ Hz, the
climbing activity of finger-like structures are quieted, allowing
persistent holes to become the dominant stable structure. This transition
from fingers to holes with increasing frequency may be an artifact
of dampened oscillation amplitude at high frequencies, reflecting
a limitation in our experimental setup.

While most of the observed phenomena have been discussed in literature
(e.g. faraday waves, holes, and fingers \cite{Merkt2004Persistent}\cite{vonKann2014Phase}),
climbing droplets have not been observed previously and are broadly
consistent with both the stress hysteresis mechanism suggested by
Deegan \cite{Deegan2010Stress} and our generalization of the same
to the notion of a ``negative viscosity'' efffect.

\subsection{Discussion and Relationship to Model Predictions\label{subsec:Discussion-and-Relationship}}

To discuss our experimental findings in relation to the modeling work
of section \ref{sec:Modeling-Results}, we will need to estimate model
parameters and review some assumptions and approximations. Given the
lack of precision in our experimental design and the weaknesses of
our modeling approach, we feel that it is only appropriate to frame
this discussion in qualitative terms and any appearance of quantitative
agreement should be understood as fortuitous and insufficient to validate
our modeling approach.

First, we will assume that the flow within a rising or falling column
of fluid can be idealized in terms of a 1D flow in a falling film,
as per sections \ref{subsec:square_gravity} and \ref{subsec:sine_gravity}.
Since the displacement field is varying in a sinusoidal pattern, we
will compare against the predictions of Figure \ref{fig:Sinusoidal droplet},
requiring estimates of the dimensionless groups $\Delta g/g_{0}$
and $\sigma_{0}/\sigma^{*}$. Because climbing is only apparent in
subfigures (c) and (d), these will be the focus of our discussion,
though we make no claims as to whether our experiments corresponded
to DST or SJ conditions.

To review, the dimensionless group $\sigma_{0}/\sigma^{*}$ compares
the wall stress to the stress for frictional contact in the case of
a vertical falling film with no oscillations. The dimensionless group
$\Delta g/g_{0}$ then compares the peak amplitude for changes in
the apparent gravitational field against the background acceleration
due to gravity. We will estimate these dimensionless groups for our
reference observations at $10$Hz and maximum volume (1cm peak-to-peak
displacements), or approximately $20$m/$\text{s}^{2}=2g_{0}$. Under
these conditions, fluid columns with approximate thickness $1$mm
were observed to climb up vertically-oriented surfaces.

To compute $\sigma_{0}/\sigma^{*}$, we will need estimates for both
$\sigma_{0}$ and $\sigma^{*}$. With respect to the former, a steady
falling film in fixed gravity gives a wall shear stress of $\sigma_{0}=\rho g_{0}h$,
where $h$ is the thickness of the fluid film in this case. From the
experiments of Merkt et. al. \cite{Merkt2004Persistent}, we see evidence
of discontinuous shear thickening from around shear stresses of $40\text{Pa}\cdot\text{s}$,
and we use this as our estimate of $\sigma^{*}$. Assuming a fluid
density of around $1.3$g/cc and a film height of $1$mm, this gives
$\sigma_{0}/\sigma^{*}\approx3$. Similarly, we estimate $\Delta g/g_{0}\approx2$.

It is worth noting that the ratio $\sigma_{0}/\sigma^{*}$ is proportional
to the film thickness, which our experiments did not directly control
for. The selection of a preferred film thickness in experiments could
reflect a ``fastest rising'' thickness but it might also simply
be constrained by capillary effects: future work would be needed to
explore this question.

In Figures \ref{fig:Sinusoidal droplet}(c) and (d), we see that model
predictions show a net climbing behavior for $\sigma_{0}/\sigma^{*}\approx3$
and $\Delta g/g_{0}\approx2$, though this does appear to lie very
near to the boundary where climbing behaviors are predicted.  In fact, the model suggests
that thicker fluid films would also show climbing
under these conditions - possibly even at a faster speed - but this was not observed. For this
reason we note that the selection of a preferred film thickness for
climbing may be more complex than what our naieve modeling approach
is prepared to consider.

Overall, we can confirm that at least one of our documented experimental
observations lies quantitatively within the regime where climbing
behaviors might be expected from our simple modeling approach. Qualitatively,
our experiments also agree with Figure\ref{fig:Sinusoidal droplet}(c)
and (d) in that a sufficienty large oscillating gravitational field
is required to induce climbing behaviors, and that droplets/films
will not climb if they are below a minimum thickness, i.e. low values
of $\sigma_{0}/\sigma^{*}$. Beyond this, our experimental design
was not sufficiently powerful or well-controlled to facilitate a more
complete comparison with vibrated fluid film predictions of Figure
\ref{fig:Sinusoidal droplet}.

\section{Summary and Conclusions}

The behavior of dense suspensions climbing up against oscillating
gravity can be described in terms of an apparent ``negative viscosity'',
as average stress and shear rate are in opposite directions.

The Wyart-Cates model for dense suspensions is used to calculate shear
rates and velocities. Average shear rates corresponding to oscillating
stresses are first calculated. As with the stress hysteresis mechanism
proposed in earlier work by Deegan, an apparent ``negative viscosity''
is only possible when the underlying flow curve exhibits discontinuous
shear thickening (DST) or shear jamming (SJ): we find no such effect
for fluids with continuous shear thickening (CST). With DST and SJ
fluids, negative viscosity effects are more readily seen when (1)
the particle fraction $\phi$ is close to random close packing $\phi_{S}$, (2) oscillations in
stress are slightly larger than the mean stress $\Delta\sigma/\sigma_0>1$, and (3) averages stresses
are comparable to the stress needed for frictional contacts to form.
Square wave forcing appears to be most efficient at exploiting a fluid's
potential for climbing, since smoothly varying forcing protocols (e.g.
sine-wave) are forced to spend more time exploring the non-inverted
portion of the flow curve at positive shear rates, where large positive
shear rates are sampled.

We believe that this apparent ``negative viscosity'' effect is likely
relevant to a wide range of phenomena observed in VVDS. As a first
line of inquiry, we provided calculations of the WC model for a falling
film under vibrating gravity. There, it was indeed found that the
discussion on oscillating shear flows was directly relevant to explaining
the conditions under which the net velocity of the film was opposed
to the average influence of gravity.

Finally, as a proof-of-principle measure, we conducted experiments
on VVDS that show fluid droplets steadily climbing up a vertically
oriented surface. These climbing droplets are a new phenomenon in
VVDS and, in our opinion, more cleanly isolate the mechanism behind
a broader range of climbing behaviors documented in the literature.
When comparing some experimental observations against our modeling
approach, we confirm that the observed climbing behavior are consistent
with a parameter space where our WC modeling framework expects a ``negative
viscosity'' effect to be observed. There are still open questions,
however, including the selection of a preferred film thickness for
climbing columns of fluid.

In future studies, and as constitutive modeling capabilities for dense
suspensions improve, it will be interesting to repeat this analysis
to account for (for example) the effects of fluid reversal and non-local
rheology. More accurate experimental measurements can also be done
to more completely characterize the phenomena of ``climbing droplets''
in VVDS. Future work may also explore a possible connection between
climbing in VVDS and the classical mechanisms of normal stress differences
and curved streamlines.

\section{Acknowledgements}

The authors would like to thank Romain Mari for sharing the results
of Figure \ref{fig:particle_sim} with us, and for allowing us to
include his work in our paper. We thank Prof Mike Cates for the support
and helpful discussions that made this project possible. We are also
grateful to an anonymous reviewer for their helpful suggestions and
thoughtful discussion, especially in the introduction and section
\ref{subsec:Discussion-and-Relationship}. Xingjian Hou thanks Dr
Michal Kwasigroch and Dr Adam Boies for their guidance on research
and funding. Work is funded by Trinity College Cambridge Projects
Fund and by the European Research Council under the Horizon 2020 Programme,
ERC grant agreement number 740269.

\printbibliography

\end{document}